\documentclass[final,1p,times]{elsarticle}

\usepackage{graphicx}

\usepackage{amsmath}
\usepackage{amssymb}
\usepackage{mathrsfs}

\usepackage[autolanguage]{numprint}

\usepackage{natbib}

\biboptions{sort&compress}

\usepackage[colorlinks]{hyperref}
\usepackage{hyphenat}

\journal{Journal of Computational Physics}

\begin{document}

\begin{frontmatter}



\title{A fast immersed boundary method for external incompressible viscous flows using lattice Green's functions}


\author{Sebastian Liska\corref{cor1}}
\ead{sliska@caltech.edu}
\author{Tim Colonius}
\ead{colonius@caltech.edu}
\cortext[cor1]{Corresponding author}
\address{Division of Engineering and Applied Science, California Institute of Technology, Pasadena, CA 91125, USA}

\begin{abstract}
A new parallel, computationally efficient immersed boundary method for solving three\hyp{}dimensional, viscous, incompressible flows on unbounded domains is presented.
Immersed surfaces with prescribed motions are generated using the interpolation and regularization operators obtained from the discrete delta function approach of the original (Peskin's) immersed boundary method.
Unlike Peskin's method, boundary forces are regarded as Lagrange multipliers that are used to satisfy the no\hyp{}slip condition.
The incompressible Navier\hyp{}Stokes equations are discretized on an unbounded staggered Cartesian grid and are solved in a finite number of operations using lattice Green's function techniques.
These techniques are used to automatically enforce the natural free-space boundary conditions and to implement a novel block-wise adaptive grid that significantly reduces the run-time cost of solutions by limiting operations to grid cells in the immediate vicinity and near-wake region of the immersed surface.
These techniques also enable the construction of practical discrete viscous integrating factors that are used in combination with specialized half-explicit Runge-Kutta schemes to accurately and efficiently solve the differential algebraic equations describing the discrete momentum equation, incompressibility constraint, and no\hyp{}slip constraint.
Linear systems of equations resulting from the time integration scheme are efficiently solved using an approximation-free nested projection technique.
The algebraic properties of the discrete operators are used to reduce projection steps to simple discrete elliptic problems, e.g. discrete Poisson problems, that are compatible with recently developed parallel fast multipole methods for difference equations.
Numerical experiments on low\hyp{}aspect\hyp{}ratio flat plates and spheres at Reynolds numbers up to $\numprint{3700}$ are used to verify the accuracy and physical fidelity of the formulation.
\end{abstract}


\end{frontmatter}

\section{Introduction}
\label{sec:intro}

Immersed boundary (IB) methods are numerical techniques for solving PDEs on Eulerian grids with immersed surfaces that are described by Lagrangian structures \cite{peskin2002,mittal2005,hou2012}.
Immersed surfaces are emulated without modifying the underlying PDE discretization by the addition of forcing terms and constraint equations resulting from the regularization of Dirac delta convolutions linking Eulerian and Lagrangian quantities.
In addition to circumventing computationally expensive body\hyp{}fitted grid generation, this approach facilitates the extensions of robust and efficient solvers, e.g. Cartesian\hyp{}grid methods, to problems involving immersed surfaces.
The original IB method \cite{peskin1972} was developed for flexible elastic structures, but has since been extended to handle more general fluid\hyp{}structure interactions, including rigid bodies and bodies with prescribed motions \cite{beyer1992,goldstein1993,glowinski1998,lai2000,taira2007,colonius2008,kallemov2015,stein2016}.
The numerous variants of the IB method and some of their higher\hyp{}order extensions are reviewed in \cite{peskin2002,mittal2005,hou2012}.
Here, we focus on \emph{distributed Lagrange multiplier (DLM) methods} \cite{glowinski1998,patankar2000,wachs2007,taira2007,colonius2008,bhalla2013,kallemov2015} since they are particularly robust IB methods for computing flows around bodies with prescribed motions \cite{hou2012}.

DLM methods treat boundary forces as Lagrange multipliers used to enforce prescribed surface boundary conditions.
For the case of fluid flows, these methods are typically expressible in forms analogous to traditional fractional\hyp{}step and projection methods and can be distinguished in part by differences in splitting errors, underlying PDEs, discretization schemes, and numerical solvers \cite{taira2007,hou2012,kallemov2015}.
The null\hyp{}space (discrete streamfunction) projection approach \cite{colonius2008} and the Rigid\hyp{}IBAMR solver \cite{kallemov2015} are examples of robust incompressible Navier\hyp{}Stokes DLM methods free of splitting errors.
The absence of splitting errors ensures that solutions retain the accuracy, stability, and physical fidelity of the PDE discretization scheme \cite{perot1993,chang2002,taira2007,colonius2008,kallemov2015}.

IB methods for external flows typically employ spatially truncated fluid domains with approximate free\hyp{}space boundary conditions, which in turn introduce \emph{blockage} errors that adversely affect the accuracy and can even change the dynamics of the numerical solution \cite{tsynkov1998,colonius2004,pradeep2004,dong2014}.
Large computational domains in combination with stretched grids \cite{yun2006, taira2007, wang2011}, local grid refinement \cite{roma1999,griffith2007,kallemov2015}, and far\hyp{}field approximation techniques \cite{colonius2008} are commonly used to reduce blockage errors.
In addition to increasing the number of computational elements, these techniques often require the use of numerical solvers that are less efficient than regular\hyp{}grid solvers (e.g. FFT techniques, multigrid, etc.) and typically result in discretization schemes that do not formally share the same conservation, commutativity, orthogonality, and symmetry properties of standard staggered Cartesian discretizations of \emph{infinite} (periodic or unbounded) domains. 

In order to eliminate the errors associated with artificial boundary conditions and to limit operations to small regions dictating the flow evolution (e.g. regions of significant vorticity), while preserving the efficiency and robustness inherent to Cartesian staggered grid methods, we proposed \cite{_liska2015} a fast incompressible Navier\hyp{}Stokes solver based on the fundamental solution, or lattice Green's function (LGF), of discrete operators.
Similar to particle and vortex methods, LGF techniques have efficient nodal distributions, automatically enforce natural free\hyp{}space boundary conditions, and can be evaluated using fast multipole methods (FMMs), e.g. the 2D serial method \cite{gillman2014} and the 3D parallel method \cite{_liska2014}.
Using the LGF\hyp{}FMM \cite{_liska2014} in combination with an projection technique that is free of splitting errors, the LGF flow solver \cite{_liska2015} computes fast, parallel solutions to the viscous integrating factor (IF) half\hyp{}explicit Runge\hyp{}Kutta (HERK) time integration scheme used to solve the velocity and pressure of the flow.

The present method numerically solves the IB formulation for the incompressible Navier\hyp{}Stokes equations given, in its continuous form, by
\begin{subequations}
  \begin{gather}
    \frac{ \partial \mathbf{u} }{ \partial t }
      + \mathbf{u} \cdot \nabla \mathbf{u}
      = - \nabla p + \frac{1}{\text{Re}} \nabla^{2} \mathbf{u}
      + \int_{\Gamma(t)} \mathbf{f}_\Gamma\left(\boldsymbol{\xi},t\right) 
      	\delta\left(\mathbf{X}\left(\boldsymbol{\xi},t\right)-\mathbf{x}\right)\,d\boldsymbol{\xi}, \label{eq:ns_ib:mom} \\
    \nabla \cdot \mathbf{u} 
      =0,\label{eq:ns_ib:cont} \\
    \int_{\mathbb{R}^3} \mathbf{u}\left(\mathbf{x},t\right) 
      	\delta\left(\mathbf{x}-\mathbf{X}\left(\boldsymbol{\xi},t\right)\right)\,d\mathbf{x}
      = \mathbf{u}_\Gamma\left(\boldsymbol{\xi},t\right), \label{eq:ns_ib:noslip}
  \end{gather}
  \label{eq:ns_ib}%
\end{subequations}
where the immersed surface $\Gamma(t)$ is parametrized by $\boldsymbol{\xi}$, and $\mathbf{X}\left(\boldsymbol{\xi},t\right)\in\Gamma(t)$.
The velocity, pressure, and Reynolds number of the flow are denoted by $\mathbf{u}\left(\mathbf{x},t\right)$, $p\left(\mathbf{x},t\right)$, and $\text{Re}$.
Here, Eq.~\eqref{eq:ns_ib:noslip} is taken to be the no\hyp{}slip condition on $\Gamma(t)$, where $\mathbf{u}_\Gamma\left(\boldsymbol{\xi},t\right) = \left[ \partial \mathbf{X} / \partial t \right]\left(\boldsymbol{\xi},t\right)$.%
  \footnote{%
  For the case of closed immersed surfaces, we limit our attention to prescribed motions that are volume conserving or, equivalently, surface velocities that satisfy the incompressibility condition $\int_{\Gamma(t)} \mathbf{u}_{\Gamma}\left(\boldsymbol{\xi},t\right) \cdot \mathbf{\hat{n}}(\boldsymbol{\xi},t)\,d\boldsymbol{\xi}=0$, where $\mathbf{\hat{n}}$ is the surface normal unit vector.}
The body force term in Eq.~\eqref{eq:ns_ib:mom}, with unknown force density $\textbf{f}_\Gamma\left(\boldsymbol{\xi},t\right)$, is computed so that $\mathbf{u}\left(\mathbf{x},t\right)$ satisfies Eq.~\eqref{eq:ns_ib:noslip}.
The fluid variables $\mathbf{u}\left(\mathbf{x},t\right)$ and $p\left(\mathbf{x},t\right)$ are defined for all $\mathbf{x}\in\mathbb{R}^3$, and subject to the boundary condition $\mathbf{u} \left( \mathbf{x}, t \right) \rightarrow 0 \,\, \text{as} \,\, \left|\mathbf{x}\right| \rightarrow \infty$.

Computationally efficient solutions for moving non-deformable (rigid) immersed surfaces are facilitated by writing Eq.~\eqref{eq:ns_ib} in an accelerating frame of reference (moving with the body), but with a change of dependent variable to the velocity in the inertial reference frame \cite{beddhu1996,kim2006,tsai2014}.
This change of variable is useful because the velocity in resulting equations tends to zero a large distances and source terms resulting from the accelerating reference frame can be absorbed into the non-linear and pressure gradient terms.
The governing equations written in the accelerating frame of reference are give by
\begin{subequations}
  \begin{gather}
    \frac{ \partial \mathbf{u} }{ \partial t }
      + ( \mathbf{u}_a \cdot \nabla ) ( \mathbf{u}_a + 2 \boldsymbol{\Omega} \times \mathbf{x}_a )
      = - \nabla q
        + \frac{1}{\text{Re}} \nabla^{2} \mathbf{u}
        + \delta \left( \Gamma(t),\mathbf{f}_\Gamma,\mathbf{x} \right)
    \label{eq:nns_ib:mom}
    \\
    \nabla \cdot \mathbf{u} 
      = 0,
    \quad
    \delta \left( \Gamma(t),\mathbf{u}, \boldsymbol{\xi} \right)
      = \mathbf{u}_{\Gamma,a}\left(\boldsymbol{\xi},t\right) + \mathbf{u}_{r}\left(\mathbf{X}\left(\boldsymbol{\xi},t\right),t\right).
  \end{gather}
  \label{eq:nns_ib}%
\end{subequations}
Here, $\mathbf{x}$ and $\mathbf{x}_a = \mathbf{x} - \mathbf{R}(t)$ denote the position vector of a point relative to the origin of the inertial- and accelerating-frame coordinates, respectively.
The accelerating-frame coordinates are taken to be centered about $\mathbf{R}(t)$, to translate with a velocity $\mathbf{U}(t)=[ \frac{d \mathbf{R}}{d t}](t)$, and to rotate about $\mathbf{R}(t)$ with an angular velocity $\boldsymbol{\Omega}(t)$ when viewed from the inertial frame.
For ease of notation, we have re-used the same symbols for the differentials as in Eq.~\eqref{eq:ns_ib}, but they now refer to the accelerating-frame coordinates, i.e. $\frac{\partial}{\partial t}$ means differentiation holding $\mathbf{x}_a$ fixed, $\nabla$ refers to the gradient in the accelerating-frame coordinates, etc.
The vectors $\mathbf{u}(\mathbf{x},t)$ and $\mathbf{u}_a(\mathbf{x},t) = \mathbf{u}(\mathbf{x},t) - \mathbf{u}_r(\mathbf{x}_a,t)$, respectively, correspond to the velocity in the inertial and accelerating reference frames, where $\mathbf{u}_r(\mathbf{x}_a,t) = \mathbf{U}(t) + \boldsymbol{\Omega}(t) \times \mathbf{x}_a$ is the velocity of a point in the accelerating frame relative to the inertial frame.
The scalar $q(\mathbf{x},t)$ is a pressure-like quantity that can be related to the inertial-frame pressure $p(\mathbf{x},t)$, up to an arbitrary time-dependent constant, using $q(\mathbf{x},t) = p(\mathbf{x},t) - \frac{1}{2} | \mathbf{u}_r(\mathbf{x}_a,t) |^2$.
Operators $\delta \left( \Gamma(t),\mathbf{f}_\Gamma, \mathbf{x} \right)$ and $\delta \left( \Gamma(t),\mathbf{u},\boldsymbol{\xi} \right)$ are shorthands for the $\delta$-function convolutions of Eq.~\eqref{eq:ns_ib:mom} and Eq.~\eqref{eq:ns_ib:noslip}.
The vectors $\mathbf{X}_a(\boldsymbol{\xi},t) = \mathbf{X}(\boldsymbol{\xi},t) - \mathbf{R}(t)$ and $\mathbf{u}_{\Gamma,a}(\boldsymbol{\xi},t) = \mathbf{u}_{\Gamma}(\boldsymbol{\xi},t) - \mathbf{u}_r(\mathbf{X}_a(\boldsymbol{\xi},t),t)$, respectively, denote the position and corresponding of velocity of a point on $\Gamma(t)$ in the accelerating reference frame.
Lastly, we clarify that the Eulerian grid and Lagrangian structure used to discretized Eq.~\eqref{eq:nns_ib} are constructed in the accelerating-frame coordinates, which implies that the Lagrangian structure of rigid surfaces can be made \emph{stationary} with respect to the Eulerian grid by specifying appropriate values for $\mathbf{R}(t)$ and $\boldsymbol{\Omega}(t)$.
This simplification is used to construct efficient solvers and pre-processing techniques that greatly reduce the run-time cost of practical flows around accelerating rigid surfaces.

In this paper, we extend the unbounded domain LGF flow solver \cite{_liska2015} to include immersed surfaces with prescribed motions using a Lagrange multiplier approach.
In Section~\ref{sec:discrete}, we discuss the discretization of Eq.~\eqref{eq:ns_ib} on unbounded fluid domains emphasizing the modifications to the LGF techniques and IF\hyp{}HERK time integration schemes of \cite{_liska2015} used to efficiently and accurately include immersed boundaries.
Linear systems of equations arising at each Runge\hyp{}Kutta stage are solved using the fast, LGF\hyp{}based, \emph{exact} nested projection technique described in Section~\ref{sec:linsys}.
Computationally expensive projection steps are shown to be reducible to simple Poisson, Poisson\hyp{}like, or viscous integrating factor problems that are compatible with the LGF\hyp{}FMM \cite{_liska2014} and make use of LGFs that are readily computed.
Significant operation count reductions for the numerical solutions of discrete elliptic equations are demonstrated for problems involving the IB regularization and interpolation operators by limiting operations to small source and target regions near the immersed surface.
Additionally, we discuss the computational considerations of some iterative and direct solution techniques for the boundary force Schur complement problem arsing in the nested projection, and demonstrate that for many practical flows around rigid surfaces a dense linear algebra pre\hyp{}processing technique results in boundary force solutions that contribute negligibly to the total run time.
In Section~\ref{sec:adapt}, we modify the block\hyp{}wise adaptive computational grid and specialize the adaptivity criteria \cite{_liska2015} to efficiently accommodate immersed surfaces.
Finally, in Section~\ref{sec:verif}, we verify the formulation through numerical experiments on flows around flat plates and spheres at Reynolds numbers up to \numprint{3700}.

\section{Discretization}
\label{sec:discrete}

\subsection{Immersed boundary method on unbounded staggered Cartesian grids}
\label{sec:discrete_grid}

In this section we highlight important features of the spatial discretization of Eq.~\ref{eq:ns_ib}.
Additional details pertaining to the flow discretization and the IB regularization/interpolation operators are provided in the discussions of the LGF flow solver \cite{_liska2015} and of the IB\hyp{}DLM methods \cite{taira2007,colonius2008,kallemov2015}, respectively.

To begin, we formally discretize Eq.~\eqref{eq:nns_ib} on an unbounded staggered Cartesian grid using second\hyp{}order finite\hyp{}volume operators,
\begin{subequations}
  \begin{gather}
    \frac{ d \mathsf{u} }{ d t }
      + \mathsf{N}( \mathsf{u}, t )
      = -\mathsf{G} \mathsf{q}
      	+ \frac{1}{\text{Re}} \mathsf{L}_\mathcal{F} \mathsf{u}
      	+ [ \mathscr{R}(t) ] \mathfrak{f},
    \label{eq:ibdnstp_raw:mom}
    \\
    \overline{\mathsf{D}} \mathsf{u} 
    	= 0,
    \quad
    [ \mathscr{I}(t) ] \mathsf{u}
      = \mathfrak{u},
  \end{gather}
  \label{eq:ibdnstp_raw}%
\end{subequations}
where $\mathsf{u}(\mathbf{n},t)$ and $\mathsf{q}(\mathbf{n},t)$ are the discrete velocity and pressure-like variables, i.e. $\mathsf{u} \approx \mathbf{u}$ and $\mathsf{q} \approx q$, at time $t\in\mathbb{R}_{\ge 0}$ and grid location $\mathbf{n}\in\mathbb{Z}^3$.
Operators $\mathsf{G}$, $\overline{\mathsf{D}}$, and $\mathsf{L}_\mathcal{F}$ are discrete gradient, divergence, and vector\hyp{}Laplace operators.
The non\hyp{}linear operator $\mathsf{N}( \mathsf{u}, t )$ is a discrete approximation of $( \mathbf{u}_a \cdot \nabla ) ( \mathbf{u}_a - 2 \boldsymbol{\Omega} \times \mathbf{x}_a )$.%
\footnote{%
  The present formulation does not assume a particular form or discretization scheme for the non-linear term $( \mathbf{u}_a \cdot \nabla ) ( \mathbf{u}_a + 2 \boldsymbol{\Omega} \times \mathbf{x}_a )$.
  For example, standard inertial-frame techniques can be used to discretize the non-linear term in its divergence form $\nabla \cdot ( \mathbf{u}_a ( \mathbf{u}_a + 2 \boldsymbol{\Omega} \times \mathbf{x}_a ) )$ \cite{beddhu1996} or in its rotational form $( \nabla \times \mathbf{u} ) \times \mathbf{u}_a + \frac{1}{2} \nabla | \mathbf{u}_a |^2$ \cite{tsai2014}.
  The numerical experiments of Section~\ref{sec:verif} are computed using the latter form and the operator stencils provided in \cite{_liska2015}.%
}
The surface functions $\mathfrak{f}(i,t)$ and $\mathfrak{u}(i,t)$ correspond to the discrete body force and surface velocity of the $i$-th Lagrangian marker located at $\mathbf{X}(\boldsymbol{\xi}_i,t)\in\Gamma(t)$, where $i\in[1,N_L]$.
We clarify that $\mathfrak{u}(i,t) = \mathbf{u}_\Gamma(\boldsymbol{\xi}_i,t)$ includes the relative velocity of the Lagrangian structure with respect to the Eulerian grid $\mathbf{u}_{\Gamma,a}(\boldsymbol{\xi}_i,t)$ (equal to zero for rigid surfaces) and the relative velocity of the accelerating-frame with respect to the inertial-frame $\mathbf{u}_r(\mathbf{X}_a(\boldsymbol{\xi}_i,t),t)$.
The time\hyp{}dependent interpolation and regularization operators $\mathscr{I}(t)$ and $\mathscr{R}(t)$ are constructed by regularizing the $\delta$-function convolutions of Eq.~\eqref{eq:ns_ib:mom} and Eq.~\eqref{eq:ns_ib:noslip}.
We limit our attention to discretizations of the form
\begin{subequations}
  \begin{gather}
    \left[[\mathscr{I}(t)]\mathsf{v}\right]^{(k)}(i,t)
      = (\Delta x)^{3} \sum_{\mathbf{n}\in\mathbb{Z}^{3}} \mathsf{v}^{(k)}(\mathbf{n},t) \delta_{\Delta x}
        \textstyle \left( \mathbf{x}^{(k)}_\mathcal{F}(\mathbf{n}) - \mathbf{X}(\boldsymbol{\xi}_i,t) \right), \\
    \left[[\mathscr{R}(t)]\mathfrak{g}\right]^{(k)}(\mathbf{n},t)
      = \sum_{i\in[1,N_L]} \mathfrak{g}^{(k)}(i,t) \delta_{\Delta x}
        \textstyle \left( \mathbf{x}^{(k)}_\mathcal{F}(\mathbf{n}) - \mathbf{X}(\boldsymbol{\xi}_i,t) \right),
  \end{gather}
\end{subequations}
where $\Delta x$ is the grid cell size, $(\cdot)^{(k)}$ denotes the $k$-th vector component, $\mathbf{x}^{(k)}_\mathcal{F}(\mathbf{n})$ is the location of the $k$-th face of the $\mathbf{n}$-th grid cell, and $\delta_{h}(\mathbf{x})=h^{-3} \prod_{k=1}^3 \phi(x_k/h)$ is a discrete delta function defined as the tensor product of the single-variable kernel function $\phi(x)$.
The operators $\mathscr{I}(t)$ and $\mathscr{R}(t)$ are adjoints (up to a scalar factor) under the standard inner product, i.e. $\mathscr{I}(t) = (\Delta x)^3 [\mathscr{R}(t)]^\dagger$.

The staggered grid consists of cells ($\mathcal{C}$) and vertices ($\mathcal{V}$) containing scalar flow quantities, and faces ($\mathcal{F}$) and edges ($\mathcal{E}$) containing vector flow quantities.
We denote real\hyp{}valued grid functions with values on $\mathcal{Q}\in\{\mathcal{C},\mathcal{F},\mathcal{E},\mathcal{V}\}$ by $\mathbb{R}^{\mathcal{Q}}$, e.g. $[\mathsf{u}](t)\in\mathbb{R}^\mathcal{F}$ and $[\mathsf{q}](t)\in\mathbb{R}^\mathcal{C}$.
Similarly, real\hyp{}valued functions with vector values specified at each Lagrangian point are denoted by $\mathbb{R}^\Gamma$, e.g. $[\mathfrak{f}](t),\,[\mathfrak{u}](t) \in \mathbb{R}^\Gamma$.
The full set of discrete vector operators used in subsequent discussions is given by %
	the discrete gradients $\mathsf{G} : \mathbb{R}^\mathcal{C} \mapsto \mathbb{R}^\mathcal{F}$ and %
		$\overline{\mathsf{G}}:\mathbb{R}^\mathcal{V} \mapsto \mathbb{R}^\mathcal{E}$, %
	the discrete curls $\mathsf{C}:\mathbb{R}^\mathcal{F} \mapsto \mathbb{R}^\mathcal{E}$ and %
		$\overline{\mathsf{C}}:\mathbb{R}^\mathcal{E} \mapsto \mathbb{R}^\mathcal{F}$, %
	the discrete divergences $\mathsf{D} : \mathbb{R}^\mathcal{E} \mapsto \mathbb{R}^\mathcal{V}$ and %
		$\overline{\mathsf{D}} : \mathbb{R}^\mathcal{F} \mapsto \mathbb{R}^\mathcal{C}$, %
	and the discrete Laplacians $\mathsf{L}_\mathcal{Q}:\mathbb{R}^\mathcal{Q} \mapsto \mathbb{R}^\mathcal{Q}$ %
		for all $\mathcal{Q}$ in $\{\mathcal{C},\mathcal{F},\mathcal{E},\mathcal{V}\}$.
The present formulation extensively makes use of the %
	symmetry (e.g. $\overline{\mathsf{D}}=-\mathsf{G}^\dagger$), %
	orthogonality (e.g. $\text{Im}( \mathsf{G} ) = \text{Null}( \mathsf{C} )$), %
	mimetic (e.g. $\mathsf{L}_\mathcal{C} = - \mathsf{G}^\dagger \mathsf{G}$, $\mathsf{L}_\mathcal{F} = - \mathsf{G} \mathsf{G}^\dagger - \mathsf{C}^\dagger \mathsf{C}$), %
	and commutativity (e.g. $\mathsf{L}_\mathcal{F} \mathsf{G} = \mathsf{G} \mathsf{L}_\mathcal{C}$) %
	properties of the discretization scheme.
Related to these properties is the fact that the scheme conserves momentum, kinetic energy, and circulation in the absence of time\hyp{}differencing errors, viscosity, and immersed surfaces provided $\mathsf{N}$ is suitably discretized \cite{morinishi1998,zhang2002}.
Under similar provisions, the adjointness of $\mathscr{R}(t):\mathbb{R}^\Gamma \mapsto \mathbb{R}^\mathcal{F}$ and $\mathscr{I}(t):\mathbb{R}^\mathcal{F} \mapsto \mathbb{R}^\Gamma$ also ensures the conservation kinetic energy \cite{peskin2002} for the case of stationary immersed surfaces.

The practical implementation of Eq.~\eqref{eq:ibdnstp_raw} is facilitated by subtracting $\frac{1}{2} \mathsf{G} \mathsf{P}(\mathsf{u}-\mathsf{u}_r)$, where $\mathsf{P}(\mathsf{v})$ is a discrete approximation of $| \mathbf{v} |^2$, from both sides of Eq.~\eqref{eq:ibdnstp_raw:mom} and by writing $\mathfrak{f}$ as $-(\Delta x)^3 \tilde{\mathfrak{f}}$.
This yields
\begin{subequations}
	\begin{gather}
    \frac{ d \mathsf{u} }{ d t }
      + \tilde{\mathsf{N}}( \mathsf{u}, t )
      = -\mathsf{G} \mathsf{d}
      	+ \frac{1}{\text{Re}} \mathsf{L}_\mathcal{F} \mathsf{u}
      	+ \left[ \mathscr{I}(t) \right]^\dagger \tilde{\mathfrak{f}},
      \label{eq:ibdnstp:mom}\\
    \overline{\mathsf{D}} \mathsf{u} 
    	= 0,
    	\label{eq:ibdnstp:cont}\\
    \left[ \mathscr{I}(t) \right] \mathsf{u}
      = \mathfrak{u},
    	\label{eq:ibdnstp:noslip}
  \end{gather}%
  \label{eq:ibdnstp}%
\end{subequations}
where $\tilde{\mathsf{N}}( \mathsf{u}, t ) = \mathsf{N}( \mathsf{u}, t ) - \frac{1}{2} \mathsf{G} \mathsf{P}(\mathsf{u}-\mathsf{u}_r)$, $\mathsf{d} = \mathsf{q} + \frac{1}{2} \mathsf{P}(\mathsf{u}-\mathsf{u}_r)$, and $\mathsf{u}_r^{(k)}(\mathbf{n},t)=\mathbf{u}^{(k)}_r(\mathbf{x}^{(k)}_\mathcal{F}(\mathbf{n}),t)$.
The non-linear term $\tilde{\mathsf{N}}( \mathsf{u}, t )$ is a discrete approximation of $( \mathbf{u}_a \cdot \nabla ) ( \mathbf{u}_a + 2 \boldsymbol{\Omega} \times \mathbf{x}_a ) - \frac{1}{2} \nabla | \mathbf{u}_a |^2 = ( \nabla \times \mathbf{u} ) \times \mathbf{u}_a$, and has the computational advantage having values that decay significantly faster at large distances compared to $\mathsf{N}( \mathsf{u}, t )$; additional details for inertial-frame flows without immersed surfaces are discussed in \cite{_liska2015}.
We call attention to the fact that the discrete equations resulting from the temporal discretizations of Eq.~\eqref{eq:ibdnstp_raw} and Eq.~\eqref{eq:ibdnstp} are, in general, different.
But, as will be shown in Section~\ref{sec:discrete_time}, the present time integration scheme evaluates $\tilde{\mathsf{N}}( \mathsf{u}, t )$ and $\mathsf{G} \mathsf{d}$ at the same times, and effectively computes the contributions from these terms as $\tilde{\mathsf{N}}( \mathsf{u}, t ) + \mathsf{G} \mathsf{d} = \mathsf{N}( \mathsf{u}, t ) + \mathsf{G} \mathsf{q}$.
This implies that the numerically integrated solutions to Eq.~\eqref{eq:ibdnstp_raw} and Eq.~\eqref{eq:ibdnstp} are \emph{equivalent} in the absence of finite precision errors.
Lastly, we note that the $\mathsf{q}(\mathbf{n},t)$ tends to an arbitrary time-dependent constant (taken to be zero) as $|\mathbf{n}| \rightarrow \infty$ and discrete pressure $\mathsf{p}(\mathbf{n},t)$ can be computed from $\mathsf{q}(\mathbf{n},t)$ using the expression $\mathsf{p} = \mathsf{q} + \frac{1}{2} \left( \mathsf{P}(\mathsf{u}-\mathsf{u}_r)- \mathsf{P}(\mathsf{u}_r) \right)$.

\subsection{Lattice Green's function technique}
\label{sec:discrete_lgf}

In this section we provide an overview of the LGF techniques \cite{_liska2014,_liska2015} and some extensions used to solve inhomogeneous, elliptic \emph{difference} equations relevant to incompressible flows on unbounded domains with immersed surfaces.
We consider the representative problem of the \emph{discrete} (7-pt) scalar Poisson equation
\begin{equation}
  [ \mathsf{L} \mathsf{x} ] (\mathbf{n})
    = \mathsf{y}(\mathbf{n}), \quad supp(\mathsf{f}) \subseteq D,
  \label{eq:dpoisson}
\end{equation}
where both $\mathsf{x}$ and $\mathsf{y}$ belong to either $\mathbb{R}^\mathcal{C}$ or $\mathbb{R}^\mathcal{V}$, and $D$ is a bounded region in $\mathbb{Z}^3$.
The procedure for solving Eq.~\eqref{eq:dpoisson} using the LGF of $\mathsf{L}$ is analogous to the procedure for solving free\hyp{}space Poisson problems using the fundamental solution of $\nabla^2$, i.e. $-1/(4 \pi |\mathbf{x}|)$.
The solution to Eq.~\eqref{eq:dpoisson} is given by the \emph{discrete} convolution
\begin{equation}
  \mathsf{u}(\mathbf{n})
    = [ \mathsf{G}_{\mathsf{L}} * \mathsf{f} ] (\mathbf{n})
    = \sum_{\mathbf{m}\in D} \mathsf{G}_{\mathsf{L}}(\mathbf{n}-\mathbf{m})
      \mathsf{f}(\mathbf{m}),
  \label{eq:dpoisson_conv}
\end{equation}
where $\mathsf{G}_{\mathsf{L}}:\mathbb{Z}^3\mapsto\mathbb{R}$ denotes the fundamental solution, or LGF, of $\mathsf{L}$ \cite{gillman2014,_liska2014}.

The present formulation computes the actions of $\mathsf{L}^{-1}$, $\mathsf{E}(\alpha)$, and $\mathsf{K}^{-1}=\left[\mathsf{E}(-\alpha)\mathsf{L}\right]^{-1}$ by evaluating expressions analogous to Eq.~\eqref{eq:dpoisson_conv} for the LGFs $\mathsf{G}_{\mathsf{L}}$, $\mathsf{G}_{\mathsf{E}}(\alpha)$, and $\mathsf{G}_{\mathsf{K}}(\alpha)$, where $\mathsf{E}(\alpha)$ is the operator exponential of $\mathsf{L}$ that is used as a viscous integrating factor in the discussion of Section~\ref{sec:discrete_time}.
Although the action of $\mathsf{K}^{-1}$ can be computed in two steps either as $[\mathsf{E}(\alpha)]\mathsf{L}^{-1}$ or $\mathsf{L}^{-1}[\mathsf{E}(\alpha)]$, significant operation count reductions are obtained by directly using $\mathsf{G}_{\mathsf{K}}(\alpha)$ to evaluate solutions to problems with source and target regions that are limited to a small neighborhood around the immersed surface, e.g. the support region of discrete delta functions.
This follows from the fact that the target and source regions of the first and second operator of either two\hyp{}step approach must be enlarged in each direction approximately by the size of the support of $\mathsf{G}_{\mathsf{E}}(\alpha)$.

The 3D LGF\hyp{}FMM \cite{_liska2014} is used by the present implementation to evaluate discrete LGF convolutions of the form given by Eq.~\eqref{eq:dpoisson_conv}.
The LGF\hyp{}FMM method is a kernel\hyp{}independent, interpolation\hyp{}based FMM for solving elliptic, constant\hyp{}coefficient difference equations on unbounded Cartesian grids to prescribed tolerances in linear algorithmic complexity.
Computational rates and parallel scaling comparable to existing fast 3D free\hyp{}space Poisson solvers have been demonstrated for the case of $\mathsf{G}_{\mathsf{L}}$ \cite{_liska2014}.
LGF\hyp{}specific computational considerations for problems involving $\mathsf{G}_{\mathsf{L}}$ and $\mathsf{G}_{\mathsf{E}}(\alpha)$ are discussed in \cite{_liska2014} and \cite{_liska2015}, respectively.
Here, we note that the fast decay of $\mathsf{G}_{\mathsf{E}}(\alpha)$ allows for the nearly identical far\hyp{}field treatment of $\mathsf{G}_{\mathsf{K}}(\alpha)$ compared to $\mathsf{G}_{\mathsf{L}}$, since for sufficiently large values $|\mathbf{n}|$ the asymptotic expansions of $\mathsf{G}_{\mathsf{L}}(\mathbf{n})$ \cite{martinsson2002,_liska2014} are also accurate approximations to $[\mathsf{G}_{\mathsf{K}}(\alpha)](\mathbf{n})$.%
\footnote{%
The value of $[\mathsf{G}_{\mathsf{E}}(\alpha)](\mathbf{n})$ decays faster than any exponential as $|\mathbf{n}|\rightarrow\infty$ for a given $\alpha \ge 0$ \cite{_liska2015}.
For typical flows, e.g. numerical experiments discussed in Section~\ref{sec:verif} and \cite{_liska2015}, $\alpha< \Delta t / \left( (\Delta x)^{2} \text{Re} \right) \lesssim 1$; for all $\alpha\in[0,1]$ and $|\mathbf{n}|>10$, the values of $\left|\left[\mathsf{G}_{\mathsf{E}}(\alpha)\right](\mathbf{n})\right|/\left|\left[\mathsf{G}_{\mathsf{E}}(\alpha)\right](\mathbf{0})\right|$ and $\left|\left[\mathsf{G}_{\mathsf{K}}(\alpha)\right](\mathbf{n})-\mathsf{G}_{\mathsf{L}}(\mathbf{n})\right|/\left|\mathsf{G}_{\mathsf{L}}(\mathbf{0})\right|$ are less than $10^{-7}$ and $10^{-9}$, respectively.}
Numerical procedures for computing $\mathsf{G}_{\mathsf{K}}(\alpha)$, and expressions in terms of Fourier and Bessel integrals for all the aforementioned LGFs are included in \ref{app:lgfs}.

\subsection{Time integration}
\label{sec:discrete_time}

Modifications to the IF\hyp{}HERK time integration technique for incompressible flows \cite{_liska2015} necessary to include immersed surfaces are discussed in this section.
We begin by considering the discrete integrating factor $\mathsf{E}_{\mathcal{Q}}( \alpha )$ corresponding to the solution operator of the discrete heat equation $d\mathsf{h}/dt = \kappa \mathsf{L}_{\mathcal{Q}} \mathsf{h}$ with $\mathsf{h}(\mathbf{n},t) \rightarrow 0$ as $|\mathbf{n}| \rightarrow \infty$.%
\footnote{%
The solution to $d\mathsf{h}/dt = \kappa \mathsf{L}_{\mathcal{Q}} \mathsf{h}$ with $\mathsf{h}(\mathbf{n},t) \rightarrow 0$ as $|\mathbf{n}| \rightarrow \infty$ is given by $\mathsf{h}(\mathbf{n},t) = \left[\mathsf{E}_{\mathcal{Q}} \left(\kappa (t-\tau)/(\Delta x)^2 \right) \mathsf{h}_\tau \right](\mathbf{n},t)$, where $\mathsf{h}_\tau(\mathbf{n})=\mathsf{h}(\mathbf{n},\tau)$.
An expression for $\mathsf{E}_{\mathcal{Q}}$ in terms of the Fourier series operator $\mathfrak{F}_\mathcal{Q}$ and the spectrum $\sigma^{\mathsf{L}}_\mathcal{Q}(\boldsymbol{\xi})$ of $(\Delta x)^2 \mathsf{L}_\mathcal{Q}$ is given by $\mathsf{E}_{\mathcal{Q}}( \alpha ) = \mathfrak{F}^{-1}_\mathcal{Q} \exp(\alpha \sigma^{\mathsf{L}}_\mathcal{Q} ) \mathfrak{F}_\mathcal{Q}$ \cite{_liska2015}.}
Taking $\mathsf{u}$ to be known at time $\tau$ and using the integrating factor $\mathsf{H}_\mathcal{Q}(t) = \mathsf{E}_{\mathcal{Q}}\left(\frac{t-\tau}{(\Delta x)^2\text{Re}}\right)$, an equivalent expression for Eq.~\eqref{eq:ibdnstp} for $t\ge\tau$ is given by
\begin{subequations}
  \begin{gather}
    \frac{ d \mathsf{v} }{ d t } 
      + \big[ \mathsf{H}_{\mathcal{F}}(t) \big]
        \tilde{\mathsf{N}} \left(
          \big[ \mathsf{H}_{\mathcal{F}}^{-1} (t) \big] \mathsf{v}, t \right)
      =
        -\mathsf{G} \mathsf{b}
          - \big[ \mathsf{H}_{\mathcal{F}} (t) \big] 
            \big[ \mathscr{I} (t) \big]^\dagger
              \tilde{\mathfrak{f}},
    \label{eq:ibdnstp_trans:mom}
    \\
    \mathsf{G}^\dagger \mathsf{v}
      = 0,
    \label{eq:ibdnstp_trans:cont}
    \\
    \big[ \mathscr{I}(t) \big] 
      \big[ \mathsf{H}_{\mathcal{F}}^{-1}(t) \big] 
        \mathsf{v}
      = \mathfrak{u},
    \label{eq:ibdnstp_trans:noslip}
  \end{gather}
  \label{eq:ibdnstp_trans}%
\end{subequations}
where $\mathsf{v} = [ \mathsf{H}_\mathcal{F}(t) ] \mathsf{u}$ and $\mathsf{b} = [ \mathsf{H}_\mathcal{C}(t) ] \mathsf{d}$.
The effect of the $\mathsf{H}_{\mathcal{F}}(t)$ and $\mathsf{H}_{\mathcal{F}}^{-1}(t)$ on the regularized forces and the no\hyp{}slip constraint \emph{cannot} be absorbed into $\tilde{\mathfrak{f}}$ and $\mathfrak{u}$ since, in general, there does \emph{not} exist an operator $\mathfrak{M}(t) : \mathcal{R}^\Gamma \mapsto \mathcal{R}^\Gamma$ such that $[ \mathsf{H}_{\mathcal{F}} (t) ] [ \mathscr{I} (t) ]^\dagger=[ \mathscr{I} (t) ]^\dagger [ \mathfrak{M}(t) ]$.
This implies that, even for the case of stationary immersed surfaces, the constraint operators explicitly depend on $t$.
The explicit temporal dependence of the constraint operators changes the character of the present system of differential algebraic equations (DAEs), i.e. Eq.~\eqref{eq:ibdnstp_trans}, compared to the analogous system of DAEs formulated in \cite{_liska2015}, i.e. Eq.~\eqref{eq:ibdnstp_trans:mom} and \eqref{eq:ibdnstp_trans:cont} with $\mathfrak{f}=0$.
As a result, the simplifications to the HERK order\hyp{}conditions for the case of trivial immersed surfaces \cite{_liska2015} need to be modified in order to develop schemes for Eq.~\eqref{eq:ibdnstp_trans} that achieve a prescribed order of accuracy.

HERK methods \cite{brasey1993,hairer1996} are used to integrate DAE systems of index 2
\begin{equation}
  \frac{dy}{dt} = f \left( y, z \right),\quad g\left( y \right) = 0,
  \label{eq:dae_brasey}
\end{equation}
where the product of partial derivatives $g_y(y) f_z(y,z)$ is non\hyp{}singular in a neighborhood about the solution, and $z$ is an unknown that must be computed so that $y$ satisfies $g(y) = 0$.
For the case of Eq.~\eqref{eq:ibdnstp_trans}, or equivalently Eq.~\eqref{eq:ibdnstp}, the operator $g_y(y) f_z(y,z)$ is invertible if and only if $\overline{\mathsf{D}}\mathsf{G}$ and $[\mathscr{I}(t)]( \mathsf{I} - \mathsf{G} (\overline{\mathsf{D}}\mathsf{G})^{-1}\overline{\mathsf{D}} )[\mathscr{I}(t)]^\dagger$ are invertible.
The invertibility of $\overline{\mathsf{D}}\mathsf{G}=\mathsf{L}_\mathcal{C}$ follows from taking $\mathsf{u}$ and $\mathsf{d}$ to tend to zero as $|\mathbf{n}| \rightarrow \infty$ \cite{_liska2015}, and the invertibility of $[\mathscr{I}(t)]\left( \mathsf{I} - \mathsf{G} \mathsf{L}_\mathcal{C}^{-1} \overline{\mathsf{D}} \right)[\mathscr{I}(t)]^\dagger$ is inferred, for practical flows, from the representative numerical experiments of Section~\ref{sec:verif} and from the discussions of similar operators arising in other IB\hyp{}DLM methods \cite{taira2007,colonius2008,kallemov2015,_goza2015}.%
\footnote{%
  The operators that arise in the discretizations \cite{taira2007,colonius2008,kallemov2015,_goza2015} are of the form $\mathsf{B} = [\mathscr{I}(t)]\mathsf{A}( \mathsf{I} - \mathsf{G} \mathsf{L}^{-1} \overline{\mathsf{D}} )[\mathscr{I}(t)]^\dagger$, where $\mathsf{A}$ is an operator resulting from to implicit treatment of the viscous term.
  Previous numerical experiments of \cite{colonius2008,kallemov2015,_goza2015} have found $\mathsf{B}$ to be well\hyp{}conditioned and solvable under grid refinement for sufficiently well\hyp{}spaced IB markers.%
}
By considering Eq.~\eqref{eq:ibdnstp_trans} in its autonomous form%
\footnote{%
  The non\hyp{}autonomous system Eq.~\eqref{eq:ibdnstp_trans} can be written as an equivalent autonomous system by taking $t$ to be part of the solution variables, e.g. $y=[\mathsf{v},t]$, and by augmenting the system of DAEs by including the trivial ODE $dt/dt=1$.%
}
 with $y=[\mathsf{v},t]$ and $z=[\mathsf{b},\mathfrak{f}]$ reveals that the corresponding partial derivatives $f_z$ and $g_y$ depend on $t$ but do not depend either $\mathsf{u}$ or $z$.
Simplifications to the general HERK order\hyp{}conditions for the special case of solely time\hyp{}dependent $f_z$ and $g_y$ are well\hyp{}described in \cite{hairer1989,brasey1993,sanderse2012}.
Tableaus and expected orders of accuracy for four representative schemes that are used to perform the numerical experiments in Section~\ref{sec:verif} are provided in \ref{app:herks}.

Next, we consider the IF\hyp{}HERK algorithm obtained using a $s$-stage HERK scheme with $\emph{shifted}$ coefficients $\tilde{a}_{i,j}$ and $\emph{shifted}$ nodes $\tilde{c}_{i}$ to integrate Eq.~\eqref{eq:ibdnstp_trans} from $t_{k}=k \Delta t$ to $t_{k+1}=(k+1) \Delta t$.
The present IF\hyp{}HERK algorithm is constructed by including the additional IB terms to the IF\hyp{}HERK algorithm of \cite{_liska2015}.
Introducing the auxiliary variables
\begin{equation}
  \mathsf{u}^{i}_{k}(\mathbf{n})
    = \left[ \mathsf{E}_\mathcal{F}
      \left(\textstyle\frac{-\tilde{c}_i \Delta t}{(\Delta x)^2\text{Re}}\right) \right]
      \mathsf{v}^{i}_{k} (\mathbf{n}),
  \,\,
  \mathsf{d}^{i}_{k}(\mathbf{n})
    = \left[ \mathsf{E}_\mathcal{F}
      \left(\textstyle\frac{-\tilde{c}_i \Delta t}{(\Delta x)^2\text{Re}}\right) \right]
      \mathsf{b}^{i}_{k}(\mathbf{n}),
  \,\, \forall i \in [1,s],
\end{equation}
and grouping the constraint variables, RHSs, and operators
\begin{equation}
  \lambda^{i}_{k} =
    \left[ \begin{array}{c}
      \mathsf{d}^{i}_{k} \\
      \tilde{\mathfrak{f}}^{i}_{k}
    \end{array} \right],
  \,\,
  \zeta^{i}_k =
    \left[ \begin{array}{c}
      0 \\
      \mathfrak{u}(t^{i}_k)
    \end{array} \right],
  \,\,
  Q^{i}_{k}
    = \left[ \begin{array}{cc}
      \mathsf{G} & [\mathscr{I}(t^{i}_k)]^\dagger
    \end{array} \right],
  \,\, \forall i \in [1,s],
\end{equation}
the $k$-th time\hyp{}step of the time integration algorithm, $\text{IF-HERK}(\mathsf{u}_{k},t_k)$, is as follows:
\begin{enumerate}
\item \emph{initialize}: set $\mathsf{u}_k^0 = \mathsf{u}_k$ and $t^{0}_{k} = t_k$.
\item \emph{multi\hyp{}stage}: for $i=1,2,\dots,s$, solve the linear system
  \begin{equation}
    \left[ \begin{array}{cc}
      (\mathsf{H}_\mathcal{F}^{i})^{-1} & Q^{i-1}_{k} \\
      (Q^{i}_{k})^\dagger & 0
    \end{array} \right]
    \left[ \begin{array}{c}
      \mathsf{u}_k^i \\
      \hat{\lambda}^{i}_{k}
    \end{array} \right]
    =
    \left[ \begin{array}{c}
      \mathsf{r}_k^i \\
      \zeta^{i}_k
    \end{array} \right],
    \label{eq:ibifherk_linsys}
  \end{equation}
  where
  \begin{equation}
    \mathsf{H}_\mathcal{F}^i =
      \mathsf{E}_\mathcal{F}\left(\textstyle\frac{ (\tilde{c}_i-\tilde{c}_{i-1}) \Delta t}{(\Delta x)^2\text{Re}}\right),
    \quad
    \mathsf{r}_k^i = \mathsf{h}_k^{i}
      + \Delta t \sum_{j=1}^{i-1} \tilde{a}_{i,j} \mathsf{w}_k^{i,j}
      + \mathsf{g}_k^{i},
    \label{eq:ibifherk_aux_1}
  \end{equation}
  \begin{equation}
    \mathsf{g}_k^i = \
      - \tilde{a}_{i,i} \Delta t\
      \tilde{\mathsf{N}}\left( \mathsf{u}_k^{i-1},t_k^{i-1}\right),
      \quad
      t_k^{i} = t_k + \tilde{c}_i \Delta t.
    \label{eq:ibifherk_g}
  \end{equation}
  Variables $\mathsf{h}_k^{i}$ and $\mathsf{w}_k^{i,j}$ are recursively computed for $i>1$ and $j>i$ using
  \begin{equation}
    \mathsf{h}_k^{i} = \mathsf{H}_\mathcal{F}^{i-1} \mathsf{h}_k^{i-1},
    \quad
    \mathsf{h}_k^{1} = \mathsf{u}_k^0
    \label{eq:ifherk_q}
  \end{equation}%
  \begin{equation}
    \mathsf{w}_k^{i,j} = \mathsf{H}_\mathcal{F}^{i-1} \mathsf{w}_k^{i-1,j},
    \quad
    \mathsf{w}_k^{i,i} = \left( \tilde{a}_{i,i} \Delta t \right)^{-1}
      \left( \mathsf{g}_k^{i} - Q^{i-1}_{k} \hat{\lambda}_k^{i} \right).
  \end{equation}
 \item \emph{finalize}: set $\mathsf{u}_{k+1} = \mathsf{u}_k^s$, $\lambda_{k+1} = \left( \tilde{a}_{s,s} \Delta t \right)^{-1} \hat{\lambda}_k^s$, and $t_{k+1} = t_k^{s}$.
\end{enumerate}
Solving Eq.~\eqref{eq:ibifherk_linsys} is expected to dominate the overall run-time cost of each IF-HERK stage, and is discussed in the next section.

\section{Fast linear system solver}
\label{sec:linsys}

\subsection{Nested projection technique}
\label{sec:linsys_proj}

In this section we demonstrate that Eq.~\eqref{eq:ibifherk_linsys} is efficiently solved using an operator\hyp{}block decomposition that is analogous to standard matrix\hyp{}block LU decompositions.
Unlike traditional projection and fractional\hyp{}step techniques \cite{chorin1967,temam1969}, which can be viewed as approximate LU decompositions \cite{perot1993}, the present approach is an \emph{exact}, i.e. free of operator approximations, projection technique \cite{chang2002}.
As a result, the method is free of ``splitting errors'' and does not make use of artificial pressure boundary conditions \cite{perot1993,chang2002,colonius2008,_liska2015}.
In contrast to 2D discrete null\hyp{}space (discrete streamfunction) methods \cite{chang2002,colonius2008}, we do not cast the discrete velocity\hyp{}pressure equations into equivalent discrete streamfunction\hyp{}vorticity equations since for 3D flows both formulation require solutions to an equal number of discrete Poisson problems but these are \emph{scalar} problems in the former and \emph{vector} problems in the latter.
The issue of which formulation is computationally faster is less obvious in the finite computational domain algorithm, discussed in Section~\ref{sec:adapt}, since the discrete velocity in the velocity\hyp{}pressure formulation is periodically ``refreshed'' from the discrete vorticity by solving a discrete \emph{vector} Poisson problem.
The arguments by the LGF flow solver \cite{_liska2015} supporting the velocity\hyp{}pressure formulation are readily extended to the present IB formulation.

We consider Eq.~\eqref{eq:ibifherk_linsys} written in terms of both Lagrange multipliers $\mathsf{d}_k^i$ and $\tilde{\mathfrak{f}}_k^i$,
\begin{equation}
M^{i}_{k}
  \left[ \begin{array}{c}
    \mathsf{u}_k^i \\
    \hat{\mathsf{d}}_k^i \\
    \hat{\mathfrak{f}}_k^i \\
  \end{array} \right]
  =
  \left[ \begin{array}{ccc}
    (\mathsf{H}_\mathcal{F}^{i})^{-1} & \mathsf{G} & (\mathscr{I}^{i-1}_{k})^\dagger \\
    \mathsf{G}^\dagger & 0 & 0 \\
    \mathscr{I}^{i}_{k} & 0 & 0
  \end{array} \right]
  \left[ \begin{array}{c}
    \mathsf{u}_k^i \\
    \hat{\mathsf{d}}_k^i \\
    \hat{\mathfrak{f}}_k^i \\
  \end{array} \right]
  =
  \left[ \begin{array}{cc}
    \mathsf{r}_k^i \\
    0 \\
    \mathfrak{u}_k^i
  \end{array} \right],
  \label{eq:ibifherk_linsys_full}
\end{equation}
where $\hat{\mathsf{d}}^{i}_{k} / \mathsf{d}_k^i = \hat{\mathfrak{f}}_k^i / \tilde{\mathfrak{f}}_k^i = \tilde{a}_{s,s}$, $\mathfrak{u}_k^i = \mathfrak{u}(t_k^i)$, and $\mathscr{I}_k^{i}=\mathscr{I}(t^{i}_k)$.
In general, $M^{i}_{k}$ is \emph{not} symmetric and cannot be symmetrized by rescaling $\hat{\mathfrak{f}}_k^i$ since the image of $(\mathscr{I}^{i-1}_{k})^\dagger$ and of $(\mathscr{I}^{i}_{k})^\dagger$ are different.
This is in contrast to similar IB methods, e.g. \cite{taira2007,colonius2008,kallemov2015}, which solve symmetric systems of equations analogous to Eq.~\eqref{eq:ibifherk_linsys}.
The asymmetry of $M^{i}_{k}$ is inherent to HERK integrations of DAE system of index 2 with time\hyp{}dependent constraint operators \cite{hairer1989,brasey1993}.%
\footnote{%
Similar asymmetries in the $(1,3)$ and $(3,1)$ operators are expected for operators analogous to $M^{i}_{k}$ arising from other standard semi\hyp{}explicit single- or multi\hyp{}step integration schemes for DAE systems of index 2, e.g. \cite{hairer1996} and references therein.
For example, the semi\hyp{}explicit two\hyp{}step Adams\hyp{}Bashforth method \cite{cao1998} solves Eq.~\eqref{eq:dae_brasey} as $y_{k+1} = y_{k} + \frac{\Delta t}{2} \left( 3 f(y_{k},z_{k}) - f(y_{k-1},z_{k-1}) \right)$, where the unknown $z_{k}$ is computed so that $0 = g(y_{k+1})$.
Here, the regularization operator $[\mathscr{I}(t_k)]^\dagger$ acting on the unknown body forces included in $f(y_{k},z_{k})$ is evaluated at an earlier time ($t$ is part of $y$) compared to the interpolation operator $\mathscr{I}(t_{k+1})$ included in $g(y_{k+1})$.}
Special cases of interest for which $M^{i}_{k}$ reduces to a symmetric operator include flows around rigid surfaces
and the $i$-th stage of HERK schemes with $\tilde{c}_{i-1}=\tilde{c}_{i}$.
Lastly, we call attention to the fact that the DAE index 2 conditions discussed in Section~\ref{sec:discrete_time} ensure the solvability of Eq.~\eqref{eq:ibifherk_linsys_full} \cite{brasey1993,fuhrer2013}, but emphasize that these conditions are satisfied only if $[\mathscr{I}(t)]\left( \mathsf{I} - \mathsf{G} \mathsf{L}_\mathcal{C}^{-1} \overline{\mathsf{D}} \right)[\mathscr{I}(t)]^\dagger$ is invertible.
The invertibility of this operator is demonstrated for a few practical flows in Section~\ref{sec:verif}.
Additionally, the invertibility of similar operators arising in other IB-DLM formulations has been discussed and numerically demonstrated for several practical flows by previous IB methods \cite{taira2007,colonius2008,kallemov2015,_goza2015}.

Solutions to Eq.~\eqref{eq:ibifherk_linsys_full} obtained from an operator\hyp{}block LU decomposition of $M^{i}_{k}$ can be written in the projection\hyp{}like form
  \begin{align}
    \begin{split}
      \mathsf{A}^{i} \mathsf{u}^* 
        &= \mathsf{r}_k^i \\
      \mathsf{B}^{i} \mathsf{d}^* 
        &= \mathsf{G}^\dagger \mathsf{u}^* \\
      \mathfrak{C}_k^{i} \mathfrak{f}^*
        &= \mathscr{I}^{i}_{k}
            [ \mathsf{u}^* - ( \mathsf{A}^{i} )^{-1}
              \mathsf{G} \mathsf{d}^* ]
           - \mathfrak{u}_k^i
    \end{split}
    \,\,
    \begin{split}
      \hat{\mathfrak{f}}_k^i 
        &= \mathfrak{f}^* \\
      \hat{\mathsf{d}}_k^i
        &= \mathsf{d}^* - ( \mathsf{B}^{i} )^{-1} 
            \mathsf{G}^\dagger ( \mathsf{A}^{i} )^{-1}
              ( \mathscr{I}^{i-1}_{k} )^{\dagger} \hat{\mathfrak{f}}_k^i\,, \\
      \mathsf{u}_k^i
        &= \mathsf{u}^* - ( \mathsf{A}^{i} )^{-1}
          [ \mathsf{G} \hat{\mathsf{d}}_k^i
            + ( \mathscr{I}^{i-1}_{k} )^{\dagger} \hat{\mathfrak{f}}_k^i ]
    \end{split}
    \label{eq:nested_proj_naive}%
  \end{align}
where 
\begin{gather}
  \mathsf{A}^{i} = (\mathsf{H}_\mathcal{F}^{i})^{-1},
  \quad
  \mathsf{B}^{i} = \mathsf{G}^\dagger ( \mathsf{A}^{i} )^{-1} \mathsf{G}, \\
  \mathfrak{C}_k^{i} = \mathscr{I}^{i}_{k} ( \mathsf{A}^{i} )^{-1}
    [ \mathsf{I}_\mathcal{F} - \mathsf{G} ( \mathsf{B}^{i} )^{-1}  \mathsf{G}^\dagger
      ( \mathsf{A}^{i} )^{-1} ]
      ( \mathscr{I}^{i-1}_{k} )^\dagger,
  \label{eq:schurs}%
\end{gather}
and $\mathsf{I}_\mathcal{F}$ is the identity operator for $\mathbb{R}^\mathcal{F}$.
Taking in account the mimetic, orthogonality, and commutativity properties of the discrete grid operators, the nested projection method Eq.~\eqref{eq:nested_proj_naive} is reduced to the more computationally convenient form
\begin{subequations}
  \begin{align}
    \mathsf{L}_\mathcal{C} \mathsf{d}^*
      &= -\mathsf{G}^\dagger \mathsf{r}_k^i \label{eq:nested_proj-ip}\\
    \mathfrak{C}_k^{i} \hat{\mathfrak{f}}_k^i
      &= \mathscr{I}^{i}_{k} \mathsf{H}_\mathcal{C}^{i}
          [ \mathsf{r}_k^i - \mathsf{G}^\dagger \mathsf{d}^* ]
        - \mathfrak{u}_k^i \label{eq:nested_proj-f}\\
    \hat{\mathsf{d}}_k^i
      &= \mathsf{d}^*
        + \mathsf{L}_\mathcal{C}^{-1} \mathsf{G}^\dagger ( \mathscr{I}^{i-1}_{k} )^{\dagger} \hat{\mathfrak{f}}_k^i \label{eq:nested_proj-pc}\\
    \mathsf{u}_k^i
      &= \mathsf{H}_\mathcal{F}^{i} [ \mathsf{r}_k^i
        - \mathsf{G} \hat{\mathsf{d}}_k^i
        - ( \mathscr{I}^{i-1}_{k} )^{\dagger} \hat{\mathfrak{f}}_k^i ] \label{eq:nested_proj-vc}.
  \end{align}
  \label{eq:nested_proj}
\end{subequations}
Similar considerations are used to reduce the \emph{force Schur complement} operator $\mathfrak{C}_k^{i}$ to the more computationally efficient form
\begin{equation}
  \mathfrak{C}_k^{i} = \mathscr{I}^{i}_{k}
    [ \mathsf{H}_\mathcal{F}^{i} + \mathsf{G} ( \mathsf{K}_\mathcal{C}^{i} )^{-1} \mathsf{G}^\dagger ]
    ( \mathscr{I}^{i-1}_{k} )^\dagger,
  \label{eq:force_schur}
\end{equation}
where $\mathsf{K}_\mathcal{C} = ( \mathsf{H}_\mathcal{C}^{i} )^{-1} \mathsf{L}_\mathcal{C}$.
As an aside, the physical interpretation of $\mathfrak{C}_k^{i}$ and its similarity to analogous operators arising in other IB methods, e.g. \cite{colonius2008,kallemov2015}, are facilitated by writing the operator as $\mathscr{I}^{i}_{k} \mathsf{H}_\mathcal{F}^{i} \mathsf{S} (\mathscr{I}^{i-1}_{k})^\dagger$, where $\mathsf{S}=\mathsf{I}_\mathcal{F} - \mathsf{G} \mathsf{L}_\mathcal{F}^{-1} \overline{\mathsf{D}} = - \overline{\mathsf{C}} \mathsf{L}_\mathcal{E}^{-1} \mathsf{C}$ is the orthogonal discrete divergence\hyp{}free projection operator.

Efficient computations of Eq.~\eqref{eq:nested_proj} make use of the flexible source and target regions of the LGF\hyp{}FMM.
This is particularly relevant to computations of $\mathfrak{C}_k^{i}$, $\mathscr{I}^{i}_{k} \mathsf{H}_\mathcal{C}^{i}$, and $\mathsf{L}_\mathcal{C}^{-1} \mathsf{G}^\dagger (\mathscr{I}^{i-1}_{k})^\dagger$ since the target and source regions can be limited to a small neighborhoods about the support of the discrete delta functions.
Since the IB markers are confined to a lower dimensional sub\hyp{}region of the overall computational grid, significant operation count reductions are expected when compared to schemes that do not limit the source and target regions of elliptic problems.%
\footnote{%
For test cases included in Section~\ref{sec:verif}, the typical computational time for Eq.~\eqref{eq:nested_proj-pc} is less than 50\% of that for Eq.~\eqref{eq:nested_proj-ip}.
Although Eq.~\eqref{eq:nested_proj-pc} typically requires significantly fewer than 50\% of the number of operations required by Eq.~\eqref{eq:nested_proj-ip}, parallel load balancing aiming to optimize Eq.~\eqref{eq:nested_proj-ip} (most computationally expensive step) results in a parallel work imbalance when computing Eq.~\eqref{eq:nested_proj-pc}.}
Formal definitions for the various sub\hyp{}regions of the adaptive block\hyp{}wise grid used in the present implementation are discussed in Section~\ref{sec:adapt_blocks}.

With the exception of $\hat{\mathfrak{f}}_k^i$, every term in Eq.~\eqref{eq:nested_proj} is efficiently computed either using the point\hyp{}operator representation of discrete operators or using the LGF\hyp{}FMM.
The remaining problem of efficient techniques for solving equations of the form $\mathfrak{C}_k^{i} \hat{\mathfrak{f}} = \mathfrak{r}$ is discussed in the following section.

\subsection{Force Schur complement solvers}
\label{sec:linsys_schur}

In this section we consider solutions to $\mathfrak{C}_k^{i} \hat{\mathfrak{f}} = \mathfrak{r}$ obtained using either iterative methods or dense linear algebra techniques.
We clarify that although the discussion of this section describes techniques for solving flows around immersed surfaces with general prescribed motions, the present implementation only considers the case of rigid surfaces.
Here, it is assumed that for asymptotically large problems the number of Lagrangian points, $N_L$, scales like $N_E^{\frac{2}{3}}$, where $N_E$ is the total number of Eulerian grid cells used in the finite computational domain.
Additionally, the action of $\mathfrak{C}_k^{i}$ is taken to be evaluated in $\mathcal{O}(N_L)$ by limiting operations of the LGF\hyp{}FMM solver to a few grid cells near the immersed surface.

We begin by considering the cases resulting in $\mathscr{I}^{i-1}_{k}=\mathscr{I}^{i}_{k}$, which include flows around rigid immersed surfaces and RK stages with $\tilde{c}_{i-1}=\tilde{c}_{i}$.
For these cases $\mathfrak{C}_k^{i}$ is symmetric positive\hyp{}definite (SPD), which makes the conjugate gradient (CG) method the natural iterative solver for $\mathfrak{C}_k^{i} \hat{\mathfrak{f}} = \mathfrak{r}$. 
This iterative method is used in \cite{colonius2008} to solve for the body forces from similar systems of equations, but, in contrast to the present technique, each iteration requires $\mathcal{O}(N_E \log N_E)$ instead of $\mathcal{O}(N_L)$.
Estimating the number of iterations to scale like $N^\frac{1}{2}_L$ \cite{colonius2008}, we expect the body forces for each RK stage to be computed in $\mathcal{O}(N_L^\frac{3}{2})$ operations, that is to say, $\mathcal{O}(N_E)$ operations.
As a result, the overall \emph{operation count} of the nested projection technique, i.e. Eq.~\ref{eq:nested_proj}, is $\mathcal{O}(N_E)$.

In order to estimate the \emph{computation time} of parallel algorithms it is necessary to account for the parallel scaling of the technique.
Similar to most parallel FMMs, the LGF\hyp{}FMM requires a minimum number of grid cells per processor, $\gamma_\text{eff}$, in order to sustain reasonable parallel efficiencies, e.g. greater than 80\%, as the number of processes, $N_p$, increases.%
\footnote{%
Here, we define the parallel efficiency as $T_{p=1} / \left( T_{p=N_p} N_p \right)$, where $T_{p=n}$ is the wall\hyp{}time of the algorithm.}
In practice, we expect an approximately constant $N_E/N_p \approx \gamma > \gamma_\text{eff}$, which implies that the action of $\mathfrak{C}_k^{i}$ is computed with $N_L/N_p \approx \alpha N_L / N_E \sim \mathcal{O}(N^{-\frac{1}{3}})$ grid cells per processor.
For sufficiently large problems, $N_L/N_p$ will be less than $\gamma_\text{eff}$ and continue to decrease as the problem size increases; thus, the CG body force solver is not expected to scale well.%
\footnote{%
Most simulations performed in Section~\ref{sec:verif} were performed with $\gamma \approx \gamma_\text{eff}$ and with $N_L/N_E < 10^{-3}$ (for fully developed wakes).
For these test cases, the parallel efficiency of evaluating $\mathfrak{C}_k^{i}$ can be estimated to be less than 10\%.}
Provided $\gamma$ is held constant, we expect the \emph{computation time} of evaluating $\mathfrak{C}_k^{i}$ to be $\mathcal{O}(N_E)$ and of the CG method be to $\mathcal{O}(N_L^\frac{1}{2} N_E)$ or, equivalently, $\mathcal{O}(N_L^2) \sim \mathcal{O}(N_E^\frac{4}{3})$.

An alternative approach is to use dense linear algebra to build and factor the matrix corresponding to $\mathfrak{C}_k^{i}$, and use its factored form to solve for $\hat{\mathfrak{f}}$.
For the case of rigid surfaces, the construction ($\mathcal{O}(N_L^2)$ operations) and factorization ($\mathcal{O}(N_L^3)$ operations) of the matrix $\mathbf{C} = \left[ \mathfrak{C}_k^{i}\right]$ only needs to be performed once per simulation as a pre\hyp{}processing step.
In fact, the factored form of $\mathbf{C}$ can be reused to compute flows that share the same geometry, RK tableau, and value of $\frac{\Delta t}{(\Delta x)^2\text{Re}}$.
Here, the Cholesky decomposition of the $\mathbf{C}$, i.e. $\mathbf{C} = \mathbf{L} \mathbf{L}^T$ where $\mathbf{L}$ is a lower\hyp{}triangular matrix, is computed in parallel using the ScaLAPACK library \cite{blackford1997}.
Backward and forward substitutions can be used to evaluate $[\hat{\mathfrak{f}}]=\mathbf{f}=\mathbf{L}^{-T}\mathbf{L}^{-1}\mathbf{r}$ in $\mathcal{O}(N_L^2)$ operations, but the inherent sequential nature of backward substitution limits the parallel speed\hyp{}up of this approach.
We circumvent this potential bottleneck by explicitly computing $\mathbf{W} = \mathbf{L}^{-1}$ ($\mathcal{O}(N_L^3)$ operations) as part of the pre\hyp{}processing step, and solve for $\mathbf{f}$ by evaluating $\mathbf{y}=\mathbf{W} \mathbf{r}$ and $\mathbf{f}=\mathbf{W}^{T} \mathbf{y}$ using parallel matrix\hyp{}vector multiplications ($\mathcal{O}(N_L^2)$ operations).%
\footnote{%
Although possible reductions, up to a factor of two, in the computation time of $\mathbf{f}$ are achieved by pre\hyp{}computing $\mathbf{W}^{T} \mathbf{W}$ and evaluating $\mathbf{f}$ using a single parallel matrix\hyp{}vector multiplication, this approach is expected to lead to a greater amplification of numerical errors compared to the two step approach described in the main text.%
}
By distributing the columns or rows of $\mathbf{W}$ and maintaining a local copy of $\mathbf{r}$, the parallel matrix\hyp{}vector multiplication is expected to achieve nearly perfect parallel efficiency for $N_p \ll N_L$, which is typical for most practical simulations.
As a point of reference, the largest problem considered in Section~\ref{sec:verif}, a sphere defined by approximately $8 \times 10^4$ IB markers, the average fraction of time spent evaluating $\hat{\mathfrak{f}}$ compared to the rest of Eq.~\eqref{eq:nested_proj} was less than 3\%.

Asymptotically, the computation time factoring $\mathbf{C}$ and inverting $\mathbf{L}$, and the memory requirements associated with storing $\mathbf{W}$ are expected to render the pre\hyp{}processing technique less efficient than the CG solver and potentially unfeasible on some computational resources.
Yet, for many practical problems, such as the test cases of Section~\ref{sec:verif}, the pre\hyp{}processing technique is expected to take a small fraction of the overall computation time and memory, and to yield significantly faster body forces solutions compared to the CG method.%
\footnote{%
For the largest simulation in Section~\ref{sec:verif}, i.e. sphere at $\text{Re}=3700$, pre\hyp{}processing only required approximately 10\% of the total computation time, with
less than 10\% of the pre\hyp{}processing time dedicated to factoring $\mathbf{C}$ and inverting $\mathbf{L}$.
For this test case, the time spent evaluating $\mathfrak{C}_k^{i}$ (roughly equal to the time a single CG iteration) is approximately 60\% of the time spent computing $\mathbf{f}=\mathbf{W}^{T}(\mathbf{W} \mathbf{r})$.
Estimating the number of CG iterations to reduce the initial residual by $\epsilon$ to be $\frac{1}{2} \sqrt{\kappa} \ln \frac{2}{\epsilon}$ \cite{shewchuk1994}, and taking $\epsilon=0.1$ (assumes a good initial guess) and $\kappa \simeq 1.1 \times10^3$ (computed condition number of $\mathbf{C}$), we expect the CG solver to require 50 iterations and, as a result, to be 30 times slower than $\mathbf{f}=\mathbf{W}^{T}(\mathbf{W}\mathbf{y})$.}

The general case of immersed surfaces with prescribed motion requires additional solution techniques since, for at least one RK stage, $\mathfrak{C}_k^{i}$ is only approximately symmetric, i.e. $\mathscr{I}^{i-1}_{k} = \mathscr{I}^{i}_{k} + \mathcal{O}(\Delta t)$.
Efficient parallel implementations of Krylov solvers such as GMRES and BiCGSTAB, and their ``flexible'' extensions \cite{saad1993,chen2012}, can be used to solve for $\hat{\mathfrak{f}}$ for the case of non\hyp{}symmetric $\mathfrak{C}_k^{i}$.%
\footnote{%
Flexible Krylov methods are often used to iteratively solve preconditioned linear systems that require additional (``inner'') iterations to approximate the action of the preconditioner.
For the present case, one possible preconditioner is the symmetrized version of $\mathfrak{C}_k^{i}$ obtained by approximating $\mathscr{I}^{i-1}_{k}$ as $\mathscr{I}^{i}_{k}$, which in turn allows for efficient ``inner'' CG iterations.}
Another approach that takes advantage of the efficiency of the CG method is to symmetrize $M_k^i$ by introducing an $\mathcal{O}(\Delta t)$ error so that $\mathscr{I}^{i-1}_{k}$ is replaced by $\mathscr{I}^{i}_{k}$.
Although this $\mathcal{O}(\Delta t)$ approximation results in solutions that still satisfy the discrete divergence\hyp{}free and non\hyp{}slip constraints, further analysis is required to determine its effect on the global (entire integration period) accuracy and stability of the solution.

\section{Adaptive computational domain}
\label{sec:adapt}

\subsection{Block\hyp{}wise adaptive grid}
\label{sec:adapt_blocks}

The present incompressible flow solver is implemented using the block\hyp{}wise adaptive grid of LGF flow solver \cite{_liska2015}.
When coupled with the LGF techniques discussed in Section~\ref{sec:discrete_lgf}, this approach has the advantage of limiting the computational domain to a small, finite region of the unbounded domain that efficiently accommodates temporally evolving solutions by dynamically adding and removing blocks.
Errors concentrated near the finite boundaries that result from neglecting values outside the finite computation domain are prevented from significantly contaminating the solution by padding the interior domain with buffer grid cells and periodically computing (``refreshing'') the algebraically\hyp{}decaying velocity perturbation from the exponentially\hyp{}decaying vorticity \cite{_liska2015},
\begin{equation}
  \mathsf{w} \leftarrow \mathsf{C} \mathsf{u},
  \quad
  \mathsf{s} \leftarrow \mathsf{L}^{-1} \mathsf{w},
  \quad
  \mathsf{u} \leftarrow \mathsf{C}^\dagger \mathsf{s}.
  \label{eq:vel_refresh}
\end{equation}
Estimates for the number of time\hyp{}steps, $Q_\text{max}$, before $\mathsf{u}$ needs to be refreshed from $\mathsf{w}$ are provided in \cite{_liska2015}, but we note here that for typical schemes multiple time\hyp{}steps, e.g. $Q_\text{max} \gtrsim 10$, can elapse before this refresh operation is required.
In the following discussion, we highlight additional key features of this approach and extend the base method \cite{_liska2015} to efficiently incorporate the immersed surfaces.

\begin{figure}[htbp]
  \begin{center}
    \includegraphics[width=\textwidth]{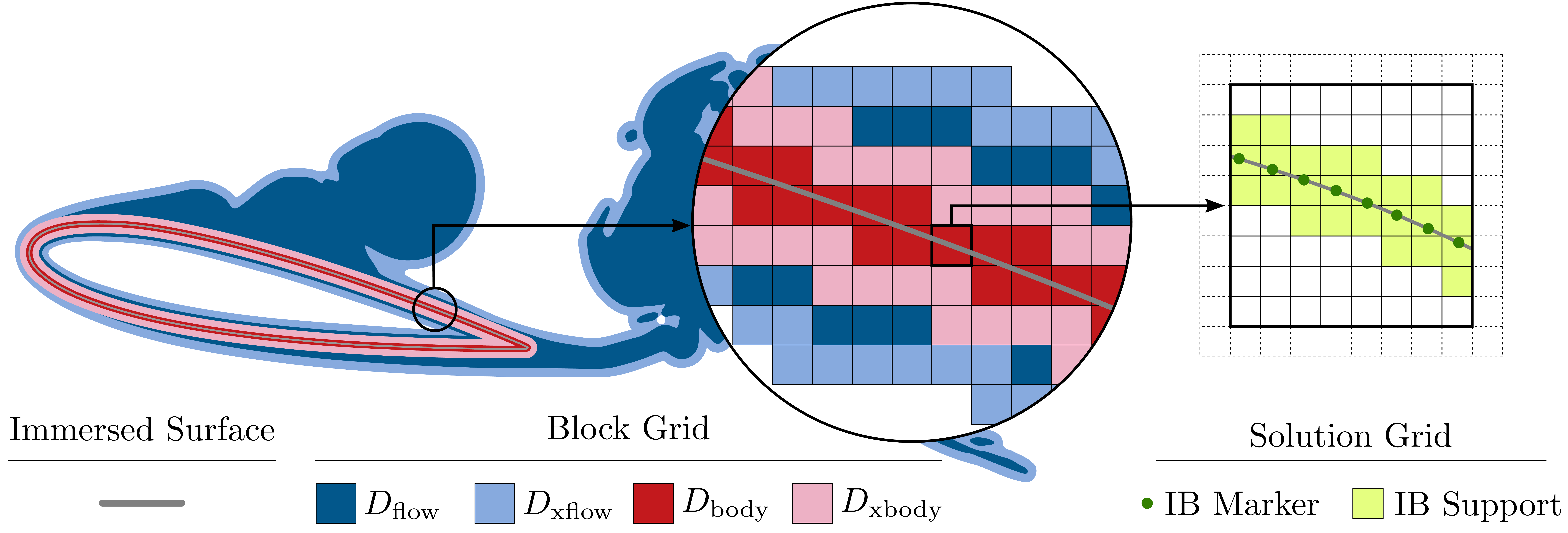}
  \end{center}
  \caption[%
    Depiction of the block-wise partitioned computational domain and finite sub-domains used by the IB-LGF method.
  ]{%
     Depiction of a 2D finite computational grid, $\emph{solution grid}$, comprised of blocks arranged on a Cartesian grid, $\emph{block grid}$.
    Distant view of nested sub\hyp{}domains $D_\text{flow} \subset D_\text{xflow}$ and $D_\text{body} \subset D_\text{xbody}$, where $D_\text{body} \subseteq D_\text{flow}$ and $D_\text{xbody} \subseteq D_\text{xbody}$ (\emph{left}).
    Zoomed\hyp{}in view illustrating the union of blocks used to define the domain (\emph{middle}).
    Magnified view of the finite staggered Cartesian grid and IB markers associated with a single block (\emph{right}).
    Dashed cells surrounding the interior cells of the isolated block correspond to ghost cells used to facilitate the parallel implementation.\label{fig:adapt_blocks}%
    }
\end{figure}

We consider partitioning the unbounded \emph{solution grid} into blocks arranged on a Cartesian \emph{block grid}.
Approximate values for each term of the IF\hyp{}HERK and the velocity refresh algorithms are computed by limiting the source (domain) and target (co\hyp{}domain) regions of discrete operators to the sub\hyp{}domains depicted in Figure~\ref{fig:adapt_blocks}.
These sub\hyp{}domains are defined as follows:
\begin{itemize}
  \item \emph{Flow domain} $D_\text{flow}$: %
  union of blocks containing non\hyp{}negligible source terms for any discrete Poisson problem solved in the present formulation, i.e. Eq.~\eqref{eq:nested_proj-ip}--\eqref{eq:nested_proj-pc} and velocity refresh.
  Flow quantities on $D_\text{flow}$ are taken to be accurate approximations of the corresponding flow quantities that would have been obtained by numerically operating on the entire unbounded domain.%
    \footnote{%
    In general, the solution can be tracked over arbitrary regions that include $D_\text{flow}$; such generalizations are discussed in \cite{_liska2015}.}
  \item \emph{Expanded flow domain} $D_\text{xflow}$: %
  union of blocks that are at most $N_\text{buf}^\text{flow}$ blocks in any direction from any block in $D_\text{flow}$.
  The value of $N_\text{buf}^\text{flow}$ is determined using the procedure discussed in \cite{_liska2015} so that the error in $D_\text{flow}$ remains below a prescribed threshold for the time\hyp{}steps between velocity refreshes.
  \item \emph{Body blocks} $D_\text{body}$: %
  union of blocks containing grid cells that are at most one grid cell away from the support of any discrete delta function during a single time\hyp{}step. %
  This implies, for example, that all non\hyp{}zero values of $\mathsf{G}^\dagger [\mathscr{I}^\dagger(t)]$ are contained within $D_\text{body}$.
  \item \emph{Expanded body domain} $D_\text{xbody}$: %
  union of blocks that are at most $N_\text{buf}^\text{body} \le N_\text{buf}^\text{flow}$ blocks in any direction from any block in $D_\text{body}$.
  We limit our attention to the case of $N_\text{buf}^\text{body} = N_\text{buf}^\text{flow} = N_\text{buf}$, and note that the subsequent discussions readily extend to the general case of $N_\text{buf}^\text{body} \ne N_\text{buf}^\text{flow}$.
\end{itemize}
Unlike domains $D_\text{flow}$ and $D_\text{xflow}$ which are recomputed only when the grid adapts \cite{_liska2015}, domains $D_\text{body}$ and $D_\text{xbody}$ are recomputed at every time\hyp{}step during which the immersed surface moves.

Summarized in Table~\ref{tab:src_trg} are the source and target regions used in the present formulation to evaluate the action of discrete operators with wide, or potentially wide, stencils (discrete kernels), i.e. $\mathsf{L}_\mathcal{Q}^{-1}$, $\mathsf{E}_\mathcal{Q}(\alpha)$, and $\mathsf{K}^{-1}_\mathcal{Q}(\alpha)$.
Similar source and target region considerations are readily deduced for all other operators, but are not discussed here since the operation count and propagation of finite boundary errors associated with these compact\hyp{}stencil operators are significantly smaller than those of the operators listed in Table~\ref{tab:src_trg}.
Also highlighted by Table~\ref{tab:src_trg} are the significant operation count reductions achieved by taking advantage of the flexible source and target regions of LGF\hyp{}FMM for cases involving immersed surfaces with $N_L \ll N_E$.

\begin{table}
  \centering
  \setlength{\tabcolsep}{4pt} 
  \begin{tabular}{rlllllll}
  Eq. & 
    \multicolumn{1}{c}{\eqref{eq:nested_proj-ip}${}^*$} &
    \multicolumn{1}{c}{\eqref{eq:nested_proj-f}${}^\dagger$} &
    \multicolumn{1}{c}{\eqref{eq:nested_proj-pc}${}^\dagger$} &
    \multicolumn{1}{c}{\eqref{eq:nested_proj-vc}} &
    \multicolumn{1}{c}{\eqref{eq:force_schur}${}^\dagger$} &
    \multicolumn{1}{c}{\eqref{eq:force_schur}${}^\dagger$} &
    \multicolumn{1}{c}{\eqref{eq:vel_refresh}} \\    
    \hline
  Operator &
    $\mathsf{L}_\mathcal{C}^{-1}$ &
    $\mathsf{H}^{i}_\mathcal{F}$ &
    $\mathsf{L}_\mathcal{C}^{-1}$ &
    $\mathsf{H}^{i}_\mathcal{F}$ &
    $(\mathsf{K}^{i}_\mathcal{F})^{-1}$ &
    $\mathsf{H}^{i}_\mathcal{F}$ &
    $\mathsf{L}_\mathcal{F}^{-1}$ \\
  Source &
    $D_\text{flow}$ &
    $D_\text{xbody}$ &
    $D_\text{body}$ &
    $D_\text{xflow}$ &
    $D_\text{body}$ &
    $D_\text{body}$ &
    $D_\text{flow}$ \\
  Target &
    $D_\text{xflow}$ &
    $D_\text{body}$ &
    $D_\text{xflow}$ &
    $D_\text{xflow}$ &
    $D_\text{body}$ &
    $D_\text{body}$ &
    $D_\text{xflow}$ \\
   Op. Count &
    $M^\mathsf{L}_{\text{f} \rightarrow \text{xf}}$ &
    $3M^\mathsf{E}_{\text{xb} \rightarrow \text{b}}$ &
    $M^\mathsf{L}_{\text{b} \rightarrow \text{xf}}$ &
    $3M^\mathsf{E}_{\text{xf} \rightarrow \text{xf}}$ &
    $M^\mathsf{K}_{\text{b} \rightarrow \text{b}}$ &
    $3M^\mathsf{E}_{\text{b} \rightarrow \text{b}}$ &
    $3M^\mathsf{L}_{\text{f} \rightarrow \text{xf}}$ \\
   Op. Scaling &
    $\mathcal{O}(N_E)$ &
    $\mathcal{O}(N_L)$ &
    $\mathcal{O}(N_E)$ &
    $\mathcal{O}(N_E)$ &
    $\mathcal{O}(N_L)$ &
    $\mathcal{O}(N_L)$ &
    $\mathcal{O}(N_E)$
  \end{tabular}
  \caption[%
    Source and target regions used by the IB-LGF method to approximate the action of non\hyp{}compact discrete operators.
  ]{%
    Source and target regions used to approximate the action of non\hyp{}compact discrete operators. %
    The number of operations required to compute the action of each operator is denoted in the second to last row; %
    operation counts for vector operations are approximated as three corresponding scalar operations.
    Superscript $*$ indicates equations originally given in the form $\mathsf{L} \mathsf{x} = \mathsf{y}$ that are written here as $\mathsf{y} = \mathsf{L}^{-1} \mathsf{x}$. %
    Superscript $\dagger$ indicate equations that are \emph{not} computed for cases without immersed surfaces.%
    \label{tab:src_trg}}
\end{table}

Temporal variations in the non\hyp{}negligible support regions of discrete operators are facilitated by adding and removing blocks to $D_\text{flow}$ and $D_\text{body}$ ($D_\text{xflow}$ and $D_\text{xbody}$ are updated accordingly).
At the end of every time\hyp{}step, flow quantities on a region that is a few grid cells greater than $D_\text{flow}$ are used to compute the block\hyp{}wise weight function $W_\text{flow}$ for each block in $D_\text{xflow}$, which in turn is used to define $D_\text{flow}$ of the next time\hyp{}step,
\begin{equation}
  D^{k+1}_\text{flow}
    = \left\{ B :
      [ W_\text{flow}(B^\prime) ] > \epsilon_\text{supp},
      \,\, B^\prime \in D^{k}_\text{xflow} \right\}.
\end{equation}
Here, we use the block\hyp{}wise weight function proposed in \cite{_liska2015},
\begin{equation}
  W_\text{flow}(B) = W(\text{pos}(B)) \max_{B \in D_\text{xflow}} \left(
    {\mu(B)}/{\mu_\text{global}},\,
    {\nu(B)}/{\nu_\text{global}} \right),
  \label{eq:weight_fun}
\end{equation}
where $W(\text{pos}(B))$ is a function of the position of block $B$,
\begin{subequations}
  \begin{alignat}{2}
    &\mu(B) = \
      \max_{\mathbf{m} \in \text{\textsl{ind}}(B)}
        | \mathsf{w}(\mathbf{n}) |, \quad
    &&\mu_\text{global} = \max_{B \in D_\text{xflow}} \mu(B), \\
    &\nu(B) = \
      \max_{\mathbf{m} \in \text{\textsl{ind}}(B)}
        | \mathsf{h}(\mathbf{n}) |, \quad
    &&\nu_\text{global} = \max_{B \in D_\text{xflow}} \nu(B).
  \end{alignat}%
\end{subequations}
Solution grid variables $\mathsf{w}=\mathsf{C}\mathsf{u}$ and $\mathsf{h}=\overline{\mathsf{D}}\tilde{\mathsf{N}}(\mathsf{u},t)$ correspond to the discrete vorticity and divergence of the Lamb vector.

For the case of $W(\mathbf{r})=1$, Eq.~\ref{eq:weight_fun} approximates the maximum normalized residual of the discrete Poisson problems Eq.~\eqref{eq:nested_proj-ip} and \eqref{eq:vel_refresh} resulting from excluding source terms outside of $D_\text{flow}$ \cite{_liska2015}.%
\footnote{%
The residuals resulting from neglecting source terms outside $D_\text{body}$ when solving the discrete Poisson problems in Eq.~\eqref{eq:nested_proj-pc} and \eqref{eq:force_schur} are zero since, by construction, $D_\text{body}$ contains all non\hyp{}zero source terms for these problems.}
Accurate solutions to the laminar\hyp{}to\hyp{}turbulence transition of thin vortex rings at Reynolds numbers up to $\numprint{7500}$ resulting in small solution grids near the ring core are reported in \cite{_liska2015} using $W(\mathbf{r})=1$.
In contrast, small solution grids are not expected to result from flows around immersed surfaces computed with $W(\mathbf{r})=1$ since vorticity is constantly generated at the surface and convected downstream.
In practice, we are often interested in accurately reproducing the flow physics in the vicinity of the immersed surface, and are willing to reduce computational costs by neglecting flow features in far\hyp{}downstream wake regions that do not significantly affect the near\hyp{}surface flow.
Point\hyp{}wise estimates for the residuals of Eq.~\eqref{eq:nested_proj-ip} and \eqref{eq:vel_refresh} based on the asymptotic $\mathcal{O}(|\mathbf{n}|^{-1})$ decay of $\mathsf{G}_\mathsf{L}$ \cite{_liska2014} indicate that errors near the immersed surface resulting from
\begin{equation}
  W(\mathbf{r}) = \frac{1}{ \max\left( \eta, \text{dist}(\mathbf{r}) \right) },
  \label{eq:weight_fun_modifer}
\end{equation}
where $\eta\ge1$ is a prescribed parameter and $\text{dist}(\mathbf{r})$ is the non\hyp{}dimensionalized distance from $\Gamma(t)$, are comparable in magnitude to those resulting from using $W(\mathbf{r}) = 1$.
Unless otherwise stated, subsequent discussions and numerical experiments take $W_\text{flow}$ to be given by Eq.~\eqref{eq:weight_fun} and \eqref{eq:weight_fun_modifer}.

\subsection{Algorithm summary}
\label{sec:adapt_summary}

In this section we summarize the present IB\hyp{}LGF method for incompressible flows on unbounded staggered Cartesian grids.
The sequence of steps performed in the IB\hyp{}LGF algorithm for an $s$-stage IF\hyp{}HERK scheme is outlined as follows:
\begin{enumerate}
  \item \emph{Pre\hyp{}processing} [Sec.~\ref{sec:linsys_schur}]: %
    (rigid surface only, optional) for each unique force Schur complement operator $\hat{\mathfrak{C}} \in \left\{ \mathfrak{C}^{i}, \forall\,i=[1,s] \right\}$, build its dense SPD matrix representation $\mathbf{C}$ by applying the operator to each standard basis vector, compute the Cholesky decomposition of $\mathbf{C}$ and invert the Cholesky factor $\mathbf{L}$ using ScaLAPACK, and store $\mathbf{W}= \mathbf{L}^{-1}$.
  \item \emph{Time integration}: for the $k$-th time\hyp{}step:
    \begin{enumerate}
      \item \emph{Grid body update} [Sec.~\ref{sec:adapt_blocks}]: %
        use the prescribed motion of $\Gamma(t)$ and the support region of discrete delta functions to compute $D_\text{body}$ and $D_\text{xbody}$ for $t\in[t_k,t_{k+1}]$.
      \item \emph{Grid flow update} [Sec.~\ref{sec:adapt_blocks}]: %
        use $D_\text{body} \subseteq D_\text{flow}$, weight function $W_\text{flow}$, threshold values $\epsilon_\text{supp}$, and $\mathsf{w}_{k} \leftarrow \mathsf{C} \mathsf{u}_{k}$ and $\mathsf{h}_{k} \leftarrow \overline{\mathsf{D}} \tilde{\mathsf{N}}(\mathsf{u}_{k},t_{k})$ to construct new $D_\text{flow}$.
        If necessary, update old $D_\text{flow}$ and $D_\text{xflow}$ by adding or removing blocks.
        Copy $\mathsf{w}_{k}$ from the old to the new solution grid and zero values on $D_\text{buffer}$.
      \item \emph{Velocity refresh} [Sec.~\ref{sec:adapt_blocks}]: %
      if either $D_\text{body}$ has been updated or the number of time\hyp{}steps since last refresh exceeds $Q_\text{max}$, compute $\mathsf{u}_{k}$ from $\mathsf{w}_k$ using Eq.~\eqref{eq:vel_refresh}.
      \item \emph{IF\hyp{}HERK} [Sec.~\ref{sec:discrete_time}]: %
        compute $\mathsf{u}_{k+1}$, ${t}_{k+1}$, $\mathsf{q}_{k+1}$, and $\mathfrak{f}_{k+1}$ using $\text{xIF-HERK}(\mathsf{u}_{k},t_k)$, where xIF\hyp{}HERK is the version of the IF\hyp{}HERK algorithm that restricts the source and target regions of discrete operators to finite sub\hyp{}regions of the solution grid, e.g. operations defined in Table~\ref{tab:src_trg}.
        Linear systems arising at each RK stage are solved using the nested projection technique Eq.~\eqref{eq:nested_proj} with an appropriate body forces solver (Section~\ref{sec:linsys_schur}).
    \end{enumerate}
\end{enumerate}

An operation count estimate for a single time\hyp{}step based on the action of all non\hyp{}compact operators, i.e. operators listed in Table~\eqref{tab:src_trg} and $\mathfrak{C}^{-1}$, is given by $M = M_{\text{flow}} + M_{\text{ib}}$, where $M_{\text{flow}}$ is the number of operations used to solve the flow without immersed surfaces \cite{_liska2015},
\begin{equation}
  M_{\text{flow}} 
    = s M^{\mathsf{L}}_{\text{f} \rightarrow \text{xf}}
    + 3 C(s) M^{\mathsf{E}}_{\text{xf} \rightarrow \text{xf}}
    + \lceil 3 M^{\mathsf{L}}_{\text{f} \rightarrow \text{xf}} \rfloor,
  \label{eq:op_count_flow}
\end{equation}
$M_{\text{ib}}$ is the number of operations required to compute the additional IB terms,
\begin{equation}
  M_{\text{ib}} 
    = s M^{\mathsf{L}}_{\text{b} \rightarrow \text{xf}}
    + s M^{\mathfrak{C}}_{\text{f} \rightarrow \text{f}}
    + 3 s M^{\mathsf{E}}_{\text{xb} \rightarrow \text{b}},
  \label{eq:op_count_ib}
\end{equation}
and $M^{\mathfrak{C}}$ is the number of operations used to solve for the body forces.
For a general $s$-stage HERK scheme with second\hyp{}order accurate constraint variables $C(s)$ is equal to $s + \left( (s-1) s \right) / 2 - 1$ \cite{_liska2015}.
The last term in Eq.~\eqref{eq:op_count_flow} is associated with the vector Poisson solve (roughly equal to three scalar Poisson solves) required by the velocity refresh procedure, which is not necessarily performed at every time\hyp{}step as indicated by the notation $\lceil\,\cdot\,\rfloor$.
All terms in Eq.~\eqref{eq:op_count_flow} and \eqref{eq:op_count_ib}, except for $M^{\mathfrak{C}}_{\text{f} \rightarrow \text{f}}$ and $M^{\mathsf{E}}_{\text{xb} \rightarrow \text{b}}$, scale as $\mathcal{O}(N_E)$.
The term $M^{\mathsf{E}}_{\text{xb} \rightarrow \text{b}}$ scales as $\mathcal{O}(N_L)$ and, for typical flows, is negligible compared to any of the $M^{\mathsf{L}}_x$ terms.
Estimates and scaling for $M^{\mathfrak{C}}_{\text{f} \rightarrow \text{f}}$ are discussed in Section~\ref{sec:linsys_schur}, but we note here that computation time spent solving for body forces is less than 3\% of the total run time for all test cases included in Section~\ref{sec:verif}.

The present MPI\hyp{}based parallel implementation partitions and distributes the support of the discrete delta functions according to the block\hyp{}wise partition and distribution of the solution grid. 
Values for all IB markers are taken to be known by all processors, which is accomplished by having values broadcast before and aggregated after the application of $\mathscr{I}^\dagger$ and $\mathscr{I}$, respectively.
Details regarding the parallel implementation of the flow solver, i.e. the IB\hyp{}LGF method without an immersed surface, are discussed in \cite{_liska2015,_liska2014}.
Load balancing is performed every time $D_\text{flow}$ or $D_\text{xflow}$ are updated.
This operation consists of optimizing the most computationally expensive operations, i.e. Eq.~\eqref{eq:nested_proj-ip} and \eqref{eq:vel_refresh}, following the procedure described in \cite{_liska2014} with the additional requirement of having all blocks belonging to $D_\text{body}$ be distributed as equally as possible across all processors.%
\footnote{
The assumption that solving Eq.~\eqref{eq:nested_proj-ip} and \eqref{eq:vel_refresh} are the most computationally expensive operations is based on the numerical experiments of Section~\ref{sec:verif} for rigid surfaces solved using the pre\hyp{}processing technique.}
Each RK stage of the representative problems of Section~\ref{sec:verif} is evaluated at a typical computation rate (normalized by the total number of MPI processes) of approximately 80 micro\hyp{}seconds per active grid cell or, equivalently, 20 micro\hyp{}seconds per active flow variable (3 velocity components and 1 pressure per grid cell).

Lastly, we clarify that the LGF\hyp{}FMM \cite{_liska2014} and the LGF flow solver \cite{_liska2015} are direct solvers that compute solutions to prescribed tolerances based on a set of parameters.
Aside from convergence criteria required for the case of iterative boundary force solutions, the IB\hyp{}LGF method does not depend on any additional parameters beyond those of the LGF flow solver \cite{_liska2015}.
Furthermore, having limited our attention to cases with $D_\text{body} \subseteq D_\text{flow}$, $D_\text{xbody} \subseteq D_\text{xflow}$, and $N^\text{body}_\text{buf} = N^\text{flow}_\text{buf}$ ensures that the procedures of LGF flow solver used to determine appropriate values for all parameters based on single threshold $\epsilon^*$ extend to the IB\hyp{}LGF method.
Subsequent discussions take $\epsilon^*$ to be equal to the grid adaptivity parameter $\epsilon_\text{supp}$.

\section{Verification examples}
\label{sec:verif}

Numerical experiments on flows around infinitely thin rectangular flat plates and spherical shells are used to verify the IB\hyp{}LGF method.
Rectangular flat plates are generated by a set of IB markers arranged in a 2D uniform Cartesian grid with a prescribed aspect\hyp{}ratio AR and angle\hyp{}of\hyp{}attack $\alpha$.
Spherical shells, subsequently referred to as spheres, are generated by placing IB markers at the centroids of the faces of an \emph{icosphere}.
Icospheres are triangulated surfaces constructed by recursively subdividing the faces and radially projecting newly created vertices onto the unit sphere of an initial icosahedron \cite{saff1997}.
The ratio of the minimum to maximum distances between any two IB markers tends to approximately 0.28 after a large number of subdivisions.

For all test cases, the minimum spacing between any two IB markers, ${\Delta s}^*$%
, is taken to be approximately equal to the grid spacing ($1.0 < {\Delta s}^*/\Delta x < 1.1$), and the smoothed version \cite{yang2009} of the 3-pt $\delta_h$ \cite{roma1999} is used to construct $\mathscr{I}$.
Boundary forces are computed using the parameter\hyp{}free Cholesky pre\hyp{}processing technique discussed in Section~\ref{sec:linsys_schur}.
Unless otherwise stated, the adaptivity threshold parameter $\epsilon^*$ is taken to be $5\times10^{-4}$, a choice that will be justified in Section~\ref{sec:verif_spheres}.
All test cases, except those for the temporal refinement studies, are performed using the HERK coefficients of Scheme A included in \ref{app:herks} and are subject to the CFL condition $|\mathbf{u}| \Delta t/\Delta x < 1$.
The time\hyp{}step size is specified so that the CFL based on the maximum point\hyp{}wise velocity remains below $0.5$ and $0.9$ for flows at $\text{Re} \le \numprint{500}$ and $\text{Re} = \numprint{3700}$, respectively, except for the first few time-steps of impulsively started flows during which the CFL is allowed to be approximately equal to unity.
We clarify that the average CFL for spheres at $\text{Re}=\numprint{3700}$ is approximately $0.6$, which is significantly lower than the large\hyp{}time maximum of $0.9$.

\subsection{Discretization error}
\label{sec:verif_error}

The convergence rates of the discretization technique is examined through spatial and temporal refinement studies of flows around spheres of radius $D$ with a prescribed velocity $\mathbf{U}(t)=\left(U_{x}(t),0,0\right)$,
\begin{equation}
	U_x(t) = \left\{%
		\begin{array}{cl}
			4 \beta U \displaystyle\int_0^t e^{ \frac{-1}{1-\left( 4 t^\prime - 1 \right)^2} } \,dt^\prime & \text{for}\,\,0 \le \frac{ t U }{ D } \le\frac{1}{2} \\
			U & \text{for}\,\, \frac{1}{2} < \frac{ t U }{ D }
		\end{array}%
	\right.,
\end{equation}
where $\beta \simeq 2.25228$ is taken so that $U_{x}(t)$ is infinitely differentiable for all $t$ (assuming $U_{x}(t)=0$ for $t<0$).
The instantaneous Reynolds number $\text{Re}=U_x(t) D / \nu$ levels to a constant equal to $100$ for $t>\frac{1}{2} D / U_\infty$.
All numerical experiments discussed in this section are performed using $\epsilon^*=10^{-8}$.

\begin{figure}[htbp]
  \begin{center}
    \includegraphics[width=\textwidth]{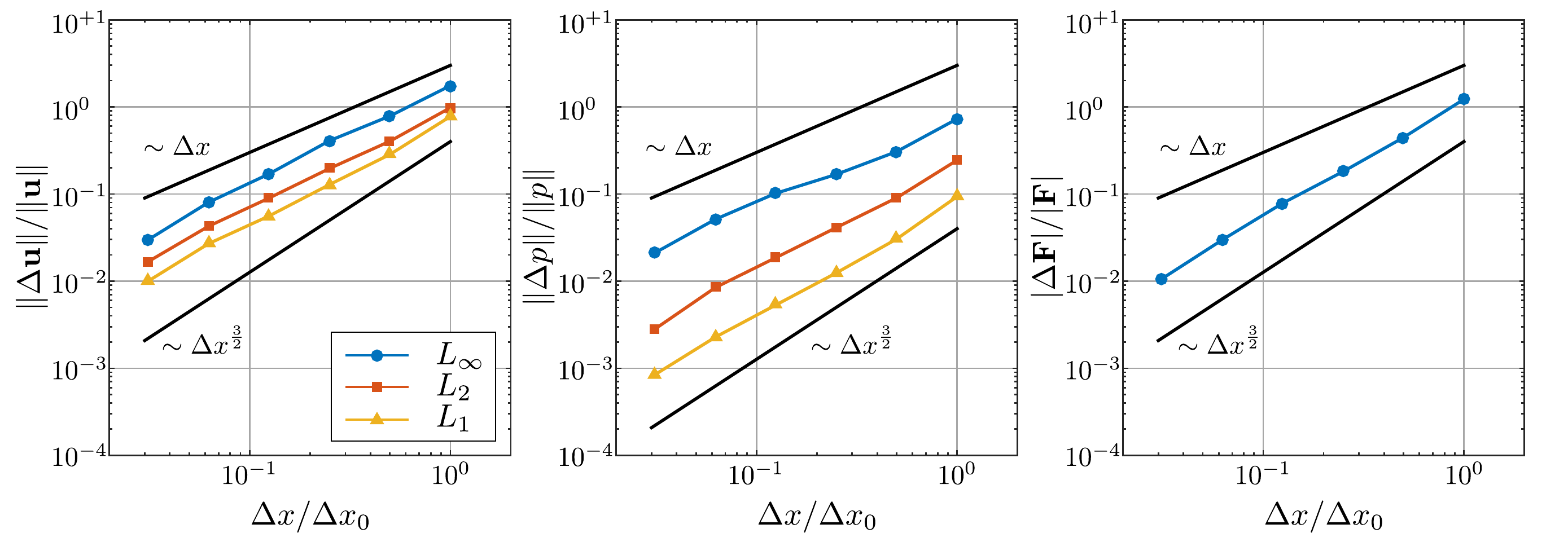}
  \end{center}
  \caption[
  Spatial convergence rates of the IB-LGF method obtained from a grid refinement study on flow around a sphere at $\text{Re}=100$.
  ]{%
  Differences in the velocity (\emph{left}), pressure (\emph{middle}), and net body force (\emph{right}) for different values of $\Delta x$ while holding $\Delta t$ fixed.
  The value of $\Delta x_0$ is equal to $\Delta x$ of coarsest test case (S.I).
  \label{fig:verif_conv_spatial}%
  }
\end{figure}

The spatial convergence study is performed using a total of seven test cases, S.I--S.VII, corresponding to spheres generated by $20 \times 4^{i-1}$ for $i=1,2,\dots,8$ IB markers.
The time\hyp{}step size, $\Delta t$, is held fixed across all test cases, and chosen so that the maximum $\text{CFL}$ for S.VII is less than $0.25$.
Estimates for the errors obtained by taking S.VII to be the reference, or \emph{true}, solution are reported in Figure~\ref{fig:verif_conv_spatial}.
As expected from analysis \cite{tornberg2004,mori2008} and numerical experiments \cite{taira2007,colonius2008,kallemov2015} of similar IB methods, the velocity $\mathbf{u} \approx \mathsf{u}$ is verified to be first\hyp{}order accurate in the $L_\infty$ norm.
Less intuitive is the fact that the pressure $p \approx \mathsf{p}$ and the net body force $\mathbf{F} \approx \sum_i \mathfrak{f}_i$ also exhibit first\hyp{}order convergence rates under the $L_\infty$ norm.

It is well\hyp{}known that the spatial smoothing inherent to the regularized delta function treatment of the immersed surfaces is unable to correctly capture the discontinuous pressure across interfaces.
The spatial regularization resulting from $\delta_h$ has been shown to lead to $\mathcal{O}(1)$ errors in the pressure near the immersed surface \cite{tornberg2004,chen2011}, which in turn prevents the $L_\infty$ convergence of the discrete pressure to the actual, continuum pressure.
The first\hyp{}order $L_\infty$ convergence rate of the pressure shown in Figure~\ref{fig:verif_conv_spatial} is a consequence of taking the reference solution to be that of S.VII as opposed to the actual solution to Eq.~\eqref{eq:ns_ib}.
Convergence rates equal or greater to first\hyp{}order follow from the fact that the continuum pressure across surfaces regularized by $\delta_h(\mathbf{x})$ is continuous and differentiable (provided a sufficiently smooth $\delta_h$) \cite{tornberg2003}.
However, for the present refinement study the convergence rate is at most first order since the regularization length\hyp{}scale $h$ is taken to be equal to $\Delta x$ of the reference (finest) solution and, as a result, will change as progressively finer test cases are considered.

Next, we consider the slightly greater than first\hyp{}order convergence rate of the body forces.
Surface stresses obtained from most IB methods are known to exhibit less than first\hyp{}order point\hyp{}wise convergence rates \cite{yang2009,kallemov2015,_goza2015} and can even grow as the immersed surface is refined \cite{_goza2015}.
Yet, it is also known that the low\hyp{}order moments for the surface stresses, such as the net force on the immersed body, are physically accurate.
This is verified by the right plot of Figure~\ref{fig:verif_conv_spatial}, which shows that the net body force, $\mathbf{F}$, convergence at a rate that is slightly greater than first\hyp{}order.
Approximate first\hyp{}order convergence rates for the net force on rigid surfaces also have been demonstrated for other IB\hyp{}DLM methods \cite{kallemov2015,_goza2015}.

Lastly, we report the $L_2$ condition number, $\kappa$, of the force Schur complement of the last RK stage, $\mathfrak{C}^s$, for each test case in Table~\ref{tab:verif_cond}.
As points of comparison, Table~\ref{tab:verif_cond} also includes values for $\kappa$ resulting from using values $\Delta s^* / \Delta x$ that are smaller ($1.00$ and $0.95$) than those used for S.I--S.VII ($1.05$).
Table~\ref{tab:verif_cond} verifies the intuitive fact, based on the finite spatial resolution of the fluid solver, that the matrix corresponding to $\mathfrak{C}^s$ rapidly becomes ill\hyp{}conditioned for values $\Delta s / \Delta x$ below certain threshold, which is approximately unity for the present test cases.
The heuristic constraint of requiring $\Delta s^* /\Delta x \ge 1$ is not universal across IB\hyp{}DLM methods; \cite{colonius2008} states that $\Delta s / \Delta x \approx 1$ results in ``reasonable'' condition numbers for $\mathfrak{C}$-like matrices computed using the 3-pt $\delta_h$ of \cite{roma1999}, and \cite{kallemov2015} numerically demonstrates that $\Delta s / \Delta x \approx 2$ results in condition numbers for $\mathfrak{C}$-like matrices computed using the 6-pt $\delta_h$ of \cite{bao2015} comparable to those listed in Table~\ref{tab:verif_cond} for the case of steady Stokes flow around a sphere.%
\footnote{%
 Condition numbers of $\mathcal{O}(10^6-10^7)$ are reported in \cite{kallemov2015} for the case of steady Stokes flow around a sphere with $\Delta s /\Delta x \approx 1$.}
We clarify that, in the continuum limit, the boundary integral operator associated with $\mathfrak{C}$ has a zero\hyp{}eigenvalue mode corresponding to the uniform compression of the sphere.
Typically, we expect that the small geometric irregularities, strict symmetry\hyp{}breaking, and slight porosity%
\footnote{%
The no-slip constraint is only enforced at a finite number of points.
The velocity at points located between IB markers is not required to satisfy the no-slip constraint.
}
of numerical immersed surfaces generated using standard discrete delta functions and well\hyp{}spaced IB\hyp{}markers, i.e. $\Delta s/ \Delta x \approx 1$, to result in non\hyp{}singular discrete $\mathfrak{C}$ operators, even if the continuous version of $\mathfrak{C}$ is singular.
Small\hyp{}eigenvalue discrete modes associated with zero\hyp{}eigenvalue continuous modes that do not satisfy the continuous divergence\hyp{}free constraint are expected to be only a small part of the solution of $\mathfrak{C} \mathfrak{f} = \mathfrak{r}$, i.e. the dot product of these modes with $\mathfrak{r}$ is small, since $\mathfrak{r}$ is interpolated from a discrete divergence\hyp{}free field.
Additional discussions and numerical examples regarding the null\hyp{}space, or lack thereof, and conditioning of $\mathfrak{C}$-like operators arising in other IB\hyp{}DLM discretizations are provided in \cite{kallemov2015,_goza2015}.
Here, the results of Table~\ref{tab:verif_cond} are used to verify the well\hyp{}posedness of Eq.~\eqref{eq:ibdnstp_trans} as a (solvable) DAE system of index 2 and to motivate the nominal value of $\Delta s^* \approx 1.05 \Delta x$ used in subsequent numerical experiments.

\begin{table}[htbp]
  \centering
  \begin{tabular}{r|rrrrrrr}
    No. IB markers &
			\multicolumn{1}{c}{$\numprint{20}$} &
			\multicolumn{1}{c}{$\numprint{80}$} &
      \multicolumn{1}{c}{$\numprint{320}$} &
      \multicolumn{1}{c}{$\numprint{1280}$} &
      \multicolumn{1}{c}{$\numprint{5120}$} &
      \multicolumn{1}{c}{$\numprint{20480}$} &
      \multicolumn{1}{c}{$\numprint{81920}$} \\ \hline
    $\Delta s^* / \Delta x \simeq 1.05$ &
      0.19 & 0.20 & 0.18 & 0.17 & 0.19 & 0.44 & 1.31 \\
    $\Delta s^* / \Delta x \simeq 1.00$ &
      0.19 & 0.52 & 0.36 & 0.42 & 1.61 & 1.41 & -- \\
    $\Delta s^* / \Delta x \simeq 0.90$ &
      2.06 & 3.13 & 1.72 & 2.69 & 8.51 & 27.94 & --
  \end{tabular}
  \caption[%
    Condition number of the force Schur complement of a sphere at $\text{Re}=100$ computed using different IB maker spacings.
  ]{%
    $L_2$ condition number of $\mathfrak{C}^s$ for different ratios of $\Delta s^* / \Delta x$.
  Values of $\Delta s^* / \Delta x \simeq 1.05$ are used in the numerical experiments S.I--S.VII.
    \label{tab:verif_cond}
  }
\end{table}

\begin{figure}[htbp]
  \begin{center}
    \includegraphics[width=\textwidth]{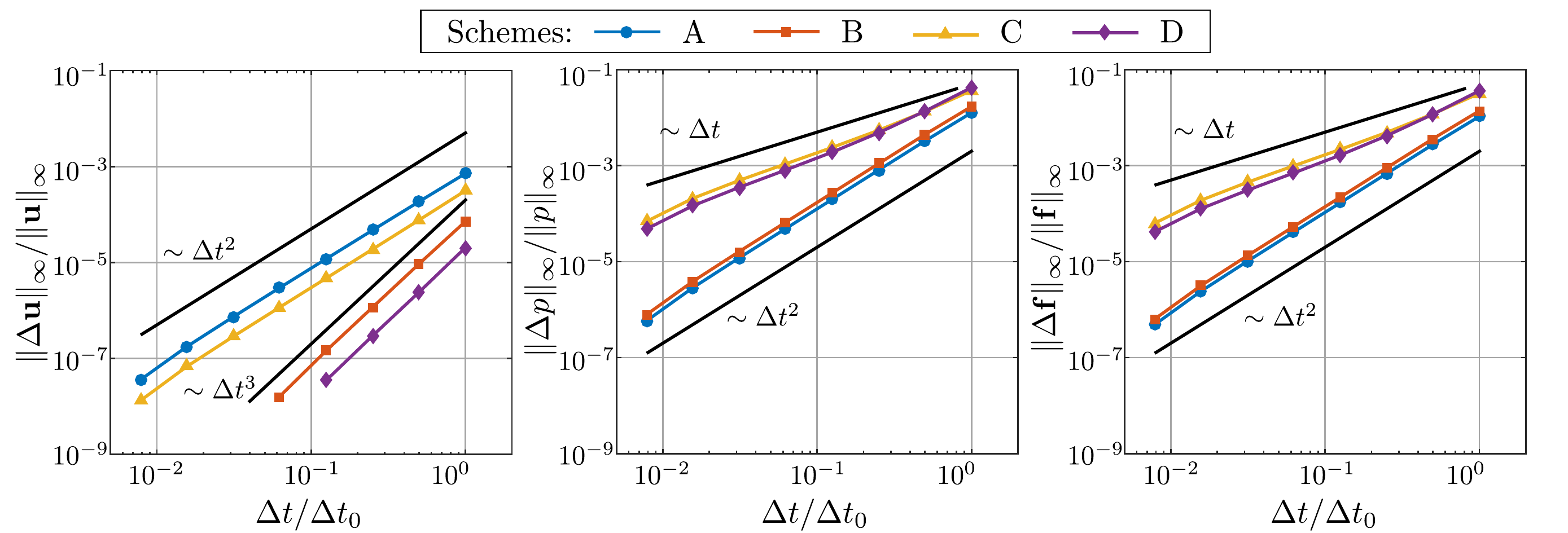}
  \end{center}
  \caption[%
    Temporal convergence rates of the IB-LGF method obtained from a grid refinement study on flow around a sphere at $\text{Re}=100$.
  ]{%
    Differences in the velocity (\emph{left}), pressure (\emph{middle}), and net body force (\emph{right}) for different values of $\Delta t$ while holding $\Delta x$ fixed.
    The value of $\Delta t_0$ is equal to $\Delta t$ of coarsest test case (T.I).
    Entries for the velocity error for Scheme B and D that are below $10^{-8}$ are excluded from the left plot since the error for these cases saturates between $10^{-9}$ and $10^{-8}$ due to prescribed adaptivity threshold $\epsilon^*=10^{-8}$.
    \label{fig:verif_conv_temporal}%
  }
\end{figure}

We now turn our attention to the temporal discretization error.
The temporal convergence studies are performed for each of the four IF\hyp{}HERK schemes included in \ref{app:herks}, Scheme A -- D, for a sphere generated by 1280 IB markers.
A total of nine test cases, T.I -- T.IX, of varying time\hyp{}step size, $\Delta t / \Delta t_0 = 2^{-i+1}$ for $i=1,2,\dots,9$, are considered for each scheme.
Here, $\Delta t_0$ is chosen such that the maximum $\text{CFL}$ of test case T.I is less than $0.5$.
Error estimates for each test case are obtained by taking T.IX of the corresponding IF\hyp{}HERK scheme to be the reference solution.%
\footnote{%
For some cases, the spatial discretization error is significantly larger than the temporal discretization error.
This does not affect the present study since the spatial discretization error is approximately the same for all test cases and our error estimates are computed as the difference of two test case solutions.}
The $L_\infty$ norm of the errors, depicted in Figure~\ref{fig:verif_conv_temporal}, verifies that the computed convergence rates of each IF\hyp{}HERK scheme with respect to $\Delta t$ is equal to the expected order of accuracy based on the HERK order conditions discussed in Section~\ref{sec:discrete_time} and \ref{app:herks}.

We emphasize that the refinement studies of this section have only verified that numerical solutions converge at the expected rate under $\Delta x$ and $\Delta t$ refinements.
The tests cases discussed in the following sections will demonstrate that the computed solutions are in fact accurate physical approximation to Eq.~\eqref{eq:ns_ib}.

\subsection{Flow around low\hyp{}aspect\hyp{}ratio rectangular plates}
\label{sec:verif-plate}

The physical fidelity of the IB\hyp{}LGF method is verified in this section by comparing solutions for impulsively\hyp{}started rectangular flat plates of chord\hyp{}length $c$ and area $A$ to previously published results.
We begin by considering the experimentally\hyp{}validated test cases \cite{taira2009} of flows around plates of $AR=2$ at $\text{Re}=100$ and $0^\circ \le \alpha \le 90^\circ$.
Here, test cases are performed taking $\Delta x =0.020c$, which is comparable to the near\hyp{}surface grid spacing of $0.025c$ used to compute these flows by other IB methods \cite{taira2009,wang2011}.
Previous refinement studies and comparisons with experimental data \cite{taira2009} indicate that the flow is sufficiently well\hyp{}resolved using the present value of $\Delta x$.

\begin{figure}[htbp]
  \begin{center}
    \includegraphics[width=0.95\textwidth]{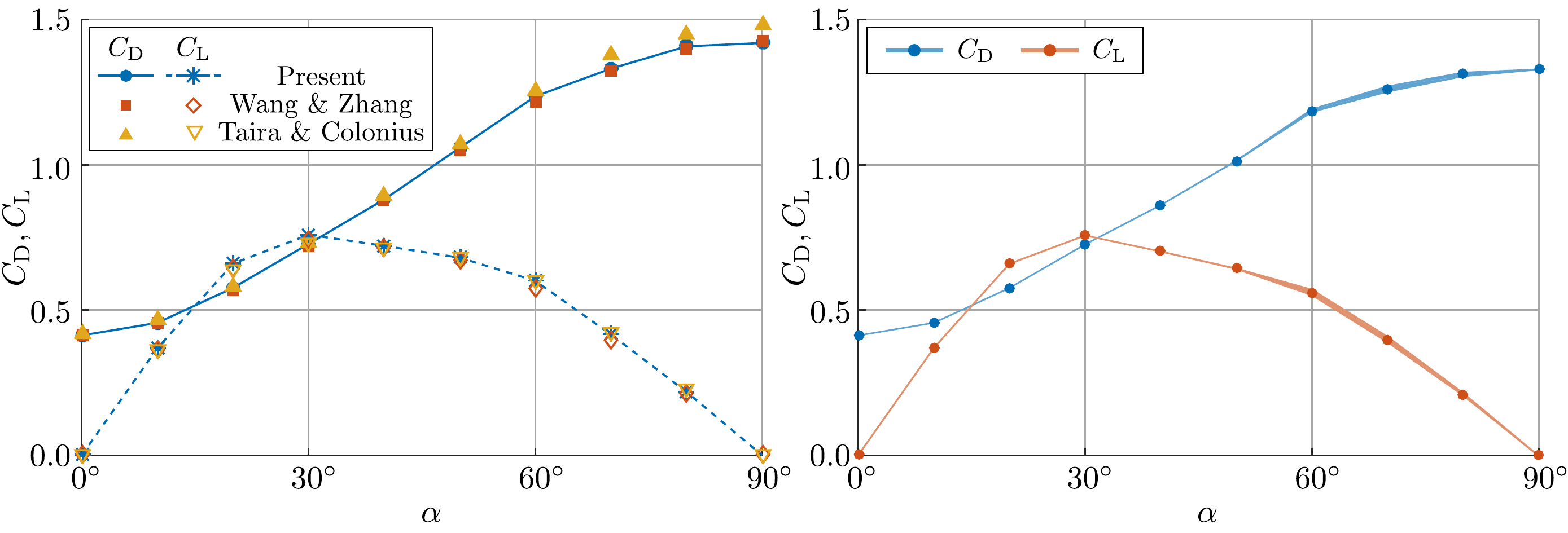}
  \end{center}
  \caption[%
    Drag and lift coefficients of rectangular flat plates of $\text{AR}=2$ at $\text{Re}=100$ and different angles\hyp{}of\hyp{}attacks.
  ]{%
    Drag and lift coefficients for rectangular flat plates of $\text{AR}=2$ at $\text{Re}=100$ and different values of $\alpha$.
    Instantaneous values at $t U / c = 13$ (\emph{left}).
    Range (shaded regions) and mean value (solid circles) of force coefficients during $50 \le t U / c \le 75$ (\emph{right}).
    \label{fig:verif_plate_lowre}%
  }
\end{figure}

The left plot of Figure~\ref{fig:verif_plate_lowre} demonstrates that the computed drag and lift coefficients, $C_\text{D}=-F_x / ( \frac{1}{2} \rho U^2 A)$ and $C_\text{L}=F_y / ( \frac{1}{2} \rho U^2 A)$, at $t U / c = 13$ are in good agreement with previously reported values \cite{taira2009,wang2011}.
The force coefficients from the three methods are nearly indistinguishable for $0^\circ \le \alpha \le 50^\circ$ and by less than 5\% for $60^\circ \le \alpha \le 90^\circ$.
The large\hyp{}time ($50 \le t U / c \le 75$) behavior of the mean and fluctuating components of $C_\text{D}$ and $C_\text{L}$ are summarized in the right plot of Figure~\ref{fig:verif_plate_lowre}.
This plot demonstrates that for $60^\circ \le \alpha < 90^\circ$ the flow is unsteady and that the large\hyp{}time mean forces are significantly different from the instantaneous forces at $t U / c = 13$.%
\footnote{%
The discussion of \cite{wang2011} regarding the present test cases states that the flow has reached a steady state at $t U / c = 13$.
A comparison between the force coefficients shown in the left and right plots of Figure~\ref{fig:verif_plate_lowre} indicates that only flows with $0^\circ \le \alpha \le 50^\circ$ have reach a steady state at $t U / c = 13$.
}
For $\alpha=60^\circ$ and $\alpha=70^\circ$ the flow is periodic with Strouhal numbers $\text{St} = F_y c \sin \alpha / U$ equal to $0.13$ and $0.11$, respectively.
In contrast, for $\alpha=80^\circ$ and $\alpha=90^\circ$ the flow is a aperiodic (at least during $50 \le t U / c \le 75$) since the force coefficients neither have a constant mean value nor a clear dominant frequency.%
\footnote{%
The value of $C_\text{L}$ for $\alpha=90^\circ$ is approximately zero for $t U / c < 10$ and decreases non\hyp{}uniformly (oscillates about local mean) to approximately $-3\times 10^{-3}$ at $t U / c = 75$.
As a point of reference, the value of $|C_\text{L}|$ for $\alpha=0^\circ$ is less than $5 \times 10^{-4}$ throughout the entire simulation period.
}
We suspect that the sensitivity of instantaneous measurements of unsteady flows to small perturbations is responsible for the larger differences across the three numerical investigations presently considered for test cases with $60^\circ \le \alpha \le 90^\circ$ when compared to the same differences for test cases with $0^\circ \le \alpha \le 50^\circ$.

Next, we consider impulsively started plates of different ARs at $\alpha=30^\circ$ and $\text{Re}=300$.
Previous numerical experiments on these flows \cite{taira2009,wang2011,wang2013} have used grid spacings of approximately $0.025c$ in the immediate vicinity of the plate (same as for the previously referenced $\text{Re}=100$ test cases).
Here, each flow is computed using (A) $\Delta x = 0.025c$ and (B) $\Delta x = 0.015 c$.

\begin{figure}[htbp]
  \begin{center}
    \includegraphics[width=0.95\textwidth]{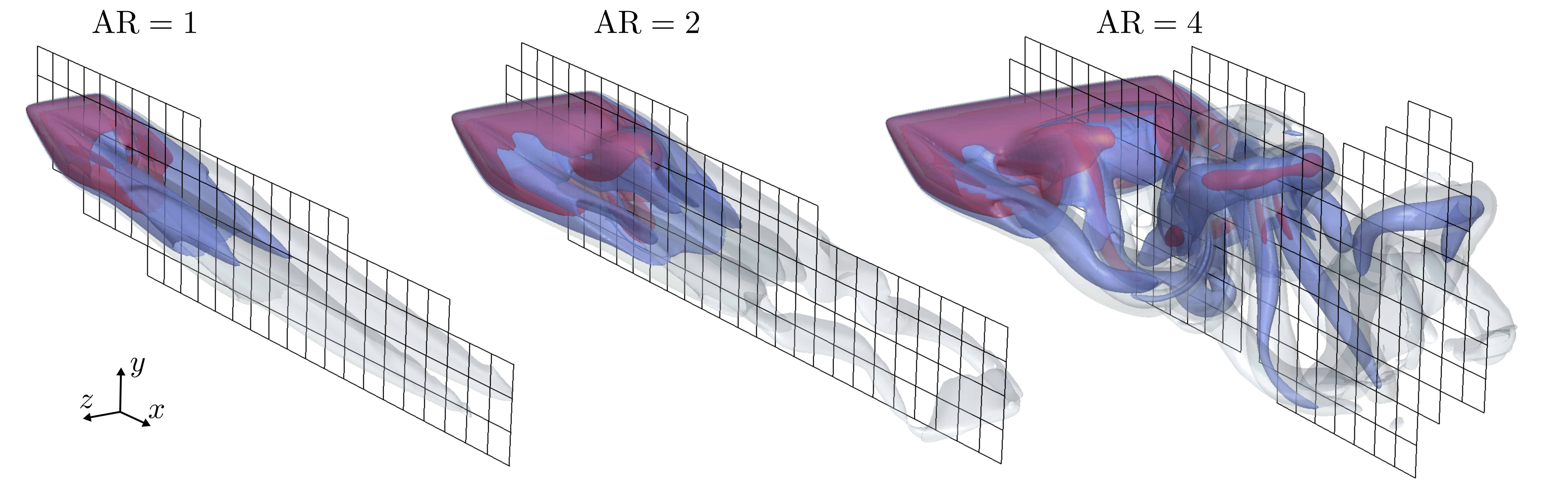}
  \end{center}
  \caption[
    Vortices in the wakes of rectangular flat plates of different aspect\hyp{}ratios at $\alpha = 30^\circ$ and $\text{Re}=300$.
  ]{%
    Vortices in the wakes of rectangular flat plates of different ARs at $\alpha = 30^\circ$ and $\text{Re}=300$.
    Shown above are iso\hyp{}surfaces of $|\boldsymbol{\omega}| c / U = 2,4,8$ at $t U / c \simeq 46.2$ computed using $\Delta x = 0.015c$.
    The union of boxes shown on the $x-y$ plane centered about the plate center depict the cross\hyp{}sectional cut of the block\hyp{}wise adaptive computational domain.
    Depicted blocks have been coarsened by a factor of two in each direction for visualization purposes.
    \label{fig:verif_plate_highre}%
  }
\end{figure}

Vortical structures in the wake of plates of $\text{AR}=1,\,2,\,\text{and}\,\,4$ are illustrated as iso\hyp{}surfaces of constant vorticity strength in Figure~\ref{fig:verif_plate_highre}.
The depicted structures are in good visual agreement with the structures reported in previous numerical experiments for $\text{AR}=1,2,4$ \cite{taira2009} and $\text{AR}=4$ \cite{wang2011,wang2013}.
Also shown in Figure~\ref{fig:verif_plate_highre} are snapshots of cross\hyp{}sectional cuts of the finite computational domains resulting from the block\hyp{}wise adaptive computational grid algorithm.
As expected from the adaptivity criteria discussed in Section~\ref{sec:adapt_blocks}, strong vortical regions near the immersed surface are efficiently tracked by adding and removing computational blocks.
For the test case of $\text{AR}=4$, the stream\hyp{}wise length of the union of blocks is approximately $[-1c,9c]$ (about the leading edge), which can be compared to the stream\hyp{}wise length $[-4c,6.1c]$ of the finite domain (with approximate boundary conditions) used in \cite{taira2009,wang2011}.

The large\hyp{}time ($50 \le t U / c \le 80$) temporal statistics for the force coefficients of the plates depicted in Figure~\ref{fig:verif_plate_highre} are included in Table~\ref{tab:plt_highre}.
Strouhal numbers of $0.12$ for $\text{AR}=4$ obtained for (A) and (B) are also reported in \cite{taira2009,wang2011}.
For all ARs, differences in mean force coefficients between (B) and \cite{taira2009,wang2011,wang2013} are less than 12\%.
The effect of the grid resolution on the accuracy of the solution can be approximated by comparing the results of the low\hyp{}resolution test cases (A) with the results of the high\hyp{}resolution test cases (B).
The values of Table~\ref{tab:plt_highre} indicate that the differences in mean force coefficients between (A) and (B) are less than 4\% for all ARs.
Since the grid resolutions of (A) and \cite{taira2009,wang2011,wang2013} are approximately the same, we suspect that modeling errors resulting from the small computational domains and approximate boundary conditions used in \cite{taira2009,wang2011,wang2013} account for most of the differences in the mean force values.%
\footnote{%
The numerical methods of \cite{taira2009} and \cite{wang2011,wang2013} use stretched and locally refined grids, respectively, to discretize computational domains of size $[-4c, 6.1c] \times [-5c, 5c] \times [-5c, 5c]$.
}

\begin{table}[htbp]
	\centering
  \small
  \setlength{\tabcolsep}{4pt}
	\begin{tabular}{rllcllclllll}
		{} &
			\multicolumn{2}{c}{$\text{AR}=1$} & {} &
			\multicolumn{2}{c}{$\text{AR}=2$} & {} &
			\multicolumn{5}{c}{$\text{AR}=4$} \\
		{} & 
			\multicolumn{1}{c}{$C_\text{D}$} & \multicolumn{1}{c}{$C_\text{L}$} & {} &
			\multicolumn{1}{c}{$\overline{C_\text{D}}$} & \multicolumn{1}{c}{$\overline{C_\text{L}}$} & {} &
			\multicolumn{1}{c}{$\overline{C_\text{D}}$} & \multicolumn{1}{c}{$\Delta C_\text{D}$} & 
				\multicolumn{1}{c}{$\overline{C_\text{L}}$} & \multicolumn{1}{c}{$\Delta C_\text{L}$} & 
					\multicolumn{1}{c}{$\text{St}$} \\
		\cline{2-3} \cline{5-6} \cline{8-12} \rule{0pt}{3ex}
		Present (A) &
    	0.543 & 0.627 & {} &
			0.504 & 0.571 & {} &
			0.620 & 0.018 & 0.791 & 0.074 & 0.118 \\
    Present (B) &
    	0.526 & 0.644 & {} &
			0.488 & 0.587 & {} &
			0.593 & 0.016 & 0.798 & 0.053 & 0.124 \\
		TC09 \cite{taira2009} &
			0.56 & 0.60 & {} &
			0.53 & 0.57 & {} &
			0.66 & -- & 0.80 & -- & 0.12 \\
		WZ11 \cite{wang2011} &
			-- & -- & {} &
			-- & -- & {} &
			-- & -- & -- & -- & 0.12 \\
		WZ13 \cite{wang2013} &
			-- & -- & {} &
			-- & -- & {} &
			-- & -- & 0.79 & -- & --
	\end{tabular}
	\caption[%
    Drag and lift coefficients, and Strouhal numbers for rectangular flat plates of different aspect\hyp{}ratios at $\alpha = 30^\circ$ and $\text{Re}=300$.
  ]{%
    Drag and lift coefficients, and Strouhal numbers for rectangular flat plates of different ARs at $\alpha = 30^\circ$ and $\text{Re}=300$.
    Present (A) and (B) correspond to test cases computed using $\Delta x / c = 2.5 \times 10^{-2}$ and $\Delta x / c = 1.5 \times 10^{-2}$, respectively.
    Results from TC09 -- \citet{taira2009}, WZ11 -- \citet{wang2011}, and WZ13 -- \citet{wang2013} are also provided.
    \label{tab:plt_highre}%
  }
\end{table}

\subsection{Flow around spheres}
\label{sec:verif_spheres}

In this section we further verify the IB\hyp{}LGF method by computing flows around impulsively started spheres.
A small perturbation $(0,\hat{u}(t),0)$ is introduced to the nominal velocity of the sphere $(U,0,0)$ in order to break axial symmetry.
We take $\hat{u}(t)$ to be non\hyp{}zero for $1 < t U / D < \frac{4}{3}$ with values equal to the bump function $ \frac{1}{10} U e^{ 1 - 1/(1-\tau^2)}$ with $\tau = 8 t - 9$.
Net body forces are reported as non\hyp{}dimensional force coefficients $C_q = F_q / ( \frac{1}{2} \rho U^2 \pi (\frac{D}{2})^2)$ for $q\in\{x,y,z\}$, and correspond to the drag ($C_\text{D}=C_x$), lateral ($C_\text{L}=C_y$), and side ($C_\text{S}=C_z$) coefficients.

First, we demonstrate that the grid adaptivity criteria and the nominal threshold value $\epsilon^* = 5 \times 10^{-4}$ accurately capture the unbounded domain flow by only tracking the solution on a small, finite region near the surface and immediate wake of a sphere at $\text{Re}=300$.
Periodic flows exhibiting planar symmetry are limited to a narrow range of Reynolds numbers that has been numerically estimated to be $280 < \text{Re} < 375$ \cite{johnson1999,mittal1999}.
The temporal periodicity and spatial symmetry about the $x$-$y$ plane of the flow \cite{johnson1999,tomboulides2000,ploumhans2002} makes this test case a challenging problem that still permits meaningful force coefficient comparisons.

\begin{figure}[htbp]
  \begin{center}
    \includegraphics[width=0.95\textwidth]{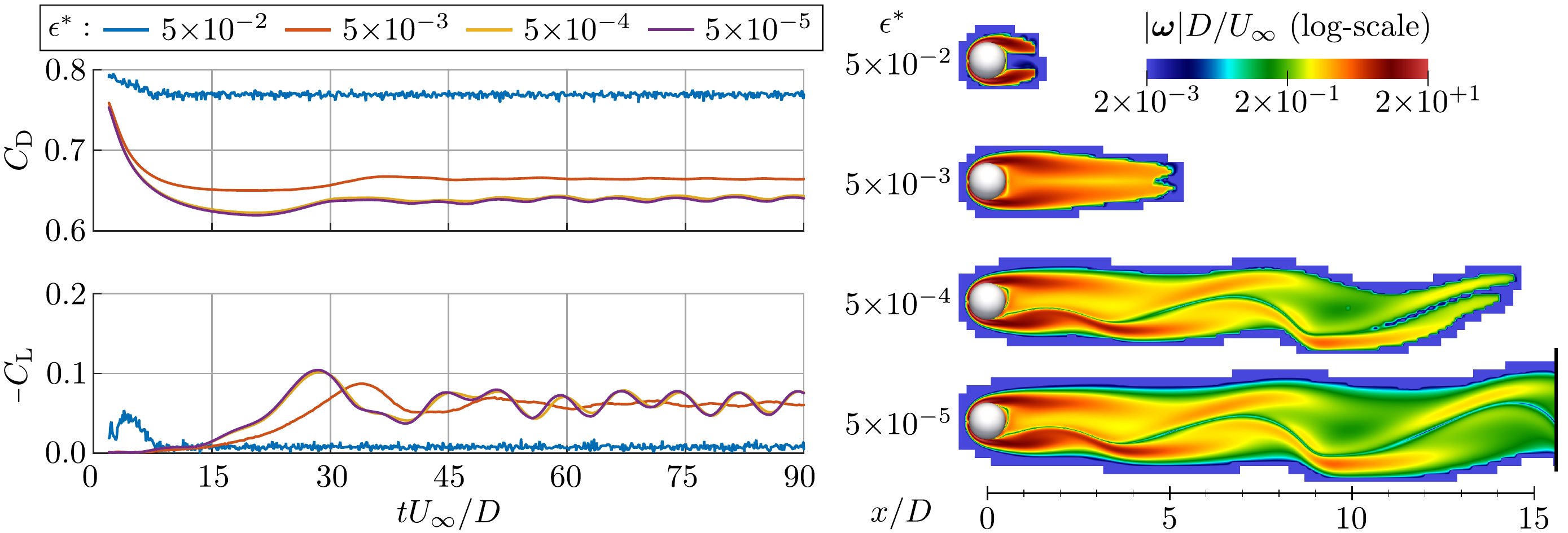}
  \end{center}
  \caption[%
    Flow around a sphere at $\text{Re}=300$ computed using different grid adaptivity threshold values.
  ]{%
    Flow around a sphere at $\text{Re}=300$ computed using different adaptive threshold, $\epsilon^*$, values.
    Time history of drag and lateral force coefficients (\emph{left}).
    Side\hyp{}view of finite computational domains at $t U / D = 80.3$ (\emph{right}).
    For each case, the finite computational domain corresponds to the non\hyp{}white region near the body and its wake.
    For the case of $\epsilon^*=5 \times 10^{-5}$ the finite computational domain continues for an additional $15D$ beyond solid black line.
    \label{fig:sph_thres}%
  }
\end{figure}

The time histories for $C_\text{D}$ and $C_\text{L}$, and snapshots of the vorticity field for spheres generated by $\numprint{20480}$ IB markers ($\Delta x \simeq 9.33 \times 10^{-3}$) computed with values of $\epsilon^* = 5 \times 10^{-i}$ for $i=2,\,3,\,4,\,\text{and}\,\,5$ are shown in Figure~\ref{fig:sph_thres}.
The maximum value of $|C_\text{S}|$ for $0 \le t U / D \le 90$ is less than $2 \times 10^{-2}$, $1 \times 10^{-2}$, $5 \times 10^{-3}$, and $2 \times 10^{-3}$, respectively, for the present test cases sorted in decreasing order of $\epsilon^*$, which in turn confirm the expect planar symmetry of the flow.
Periodic oscillations in the time history of force coefficients are clearly observed for $\epsilon^* \le 5 \times 10^{-4}$, but not for $\epsilon^* \ge 5 \times 10^{-3}$.
The apparent stabilization of the flow for $\epsilon^* \ge 5 \times 10^{-3}$ is expected from the fact that the finite computational domains of these test cases do not support a complete wavelength of the wake instability.
In contrast, the computational domain of $\epsilon^* = 5 \times 10^{-4}$ supports at least one full wavelength of the instability and is able to accurately reproduce the large\hyp{}time mean and oscillatory components of $C_\text{D}$ and $C_\text{L}$ obtained by the most conservative test case, i.e. $\epsilon^*=5 \times 10^{-5}$.

Quantitative estimates for the errors resulting from non\hyp{}zero values of $\epsilon^*$ are computed here as the differences in the force coefficient $C$ between two test cases:
	\begin{equation}
		\varepsilon^{I}(C) = \textstyle \max_{t \in T_A}{\left|C(t)-C^\prime(t)\right|},
    \,\,
		\varepsilon^{II}_\text{\textsc{ext}}(C) = \left| \text{\textsc{ext}}_{t \in T_B}{C(t)} - \text{\textsc{ext}}_{t \in T_B}{C^\prime(t)}\right|,
    \label{eq:thres_errors}
	\end{equation}
where $T_A U /D = [2,30]$, $T_B U /D = [75,90]$, and $\text{\textsc{ext}}$ is either the minimum ($\min$) or maximum ($\max$) extremum.
Error estimates obtained by taking the force coefficients of $\epsilon^*=5 \times 10^{-5}$ to be the reference values are provided in Table~\ref{tab:sphere_thres}.
As expected, the error associated with neglecting values outside $D_\text{flow}$ based on the criteria of Section~\ref{sec:adapt_blocks} is approximately proportional to $\epsilon^*$.
The results of Table~\ref{tab:sphere_thres} indicate that, in the absence of discretization errors, the forces computed using the nominal threshold, $\epsilon^*=5 \times 10^{-4}$, are accurate up to 0.6\% of the actual physical forces.%
\footnote{%
Error percentage is computed as the maximum value obtained after normalizing the errors in $C_\text{D}$ and $C_\text{L}$ provided in Table~\ref{tab:sphere_thres} by the large\hyp{}time total force coefficient $C_\text{T}=|\mathbf{F}|/( \frac{1}{2} \rho U^2 \pi (\frac{D}{2})^2) \simeq 0.66$.
}

\begin{table}[htbp]
	\centering
	\begin{tabular}{c|ccc|ccc}
		\multicolumn{1}{c|}{$\epsilon^*$} &
		\multicolumn{1}{c}{$\varepsilon^{I}(C_\text{D})$} &
		\multicolumn{1}{c}{$\varepsilon^{II}_\text{min}(C_\text{D})$} &
		\multicolumn{1}{c|}{$\varepsilon^{II}_\text{max}(C_\text{D})$} &
		\multicolumn{1}{c}{$\varepsilon^{I}(C_\text{L})$} &
		\multicolumn{1}{c}{$\varepsilon^{II}_\text{min}(C_\text{L})$} &
		\multicolumn{1}{c}{$\varepsilon^{II}_\text{max}(C_\text{L})$} \\
		\hline
		\rule{0pt}{3ex}%
		$5 \times 10^{-2}$ & 0.1600 & 0.1117 & 0.1386 & 0.1038 & 0.0470 & 0.0522 \\
		$5 \times 10^{-3}$ & 0.0311 & 0.0269 & 0.0235 & 0.0509 & 0.0108 & 0.0116 \\
		$5 \times 10^{-4}$ & 0.0030 & 0.0027 & 0.0022 & 0.0041 & 0.0003 & 0.0002
	\end{tabular}
	\caption[%
    Estimates for the error resulting from non\hyp{}zero grid adaptivity threshold values for a sphere at $\text{Re}=300$.
  ]{%
    Estimates for the error resulting from non\hyp{}zero values of $\epsilon^*$ for a sphere at $\text{Re}=300$.
    Reference values, i.e. $C^\prime$ in Eq.~\eqref{eq:thres_errors}, are taken from test case with $\epsilon^* = 5 \times 10^{-5}$.
    \label{tab:sphere_thres}%
  }
\end{table}

Having verified the error of the adaptive grid, we now turn our attention to confirming the physical fidelity of the IB\hyp{}LGF method by comparing with previous investigations of flow around spheres.
Large\hyp{}time ($90 \le t U /D \le 150$) force statistics of spheres computed at $\text{Re}=100,\,200,\, \text{and}\,\, 300$ are compared in Table~\ref{tab:sphere_coeffs}.
The flow is steady and axi\hyp{}symmetric at $\text{Re}=100$, and steady and planar $x$-$y$ symmetric at $\text{Re}=250$.
At $\text{Re}=250$ and $\text{Re}=300$ the absence of axial symmetry results in a non\hyp{}zero $C_\text{L}$.
Values computed using a coarse ($\numprint{5120}$ IB markers and $\Delta x \simeq 1.8 \times 10^{-2}$) and a fine ($\numprint{20480}$ IB markers and $\Delta x \simeq  8.8 \times 10^{-3}$) grid are shown by Table~\ref{tab:sphere_coeffs} to be consistent with the range of previously reported values.

\begin{table}[htbp]
	\centering
  \small
  \setlength{\tabcolsep}{4pt}
	\begin{tabular}{rlllcllclllll}
		{} &
			\multicolumn{3}{c}{$\text{Re}=100$} & {} &
			\multicolumn{2}{c}{$\text{Re}=250$} & {} &
			\multicolumn{5}{c}{$\text{Re}=300$} \\
		{} & 
			{} & \multicolumn{1}{c}{$C_\text{D}$} & {} & {} &
			\multicolumn{1}{c}{$C_\text{D}$} & \multicolumn{1}{c}{$C_\text{L}$} & {} &
			\multicolumn{1}{c}{$\overline{C_\text{D}}$} & \multicolumn{1}{c}{$\Delta C_\text{D}$} & 
				\multicolumn{1}{c}{$\overline{C_\text{L}}$} & \multicolumn{1}{c}{$\Delta C_\text{L}$} & 
					\multicolumn{1}{c}{$\text{St}$} \\
		\cline{3-3} \cline{6-7} \cline{9-13} \rule{0pt}{3ex}
    Present (A) &
    	{} & 1.084 & {} & {} &
			0.709 & 0.060 & {} &
			0.665 & 0.0024 & 0.067 & 0.013 & 0.133 \\
    Present (B) &
    	{} & 1.086 & {} & {} &
			0.694 & 0.059 & {} &
			0.656 & 0.0024 & 0.065 & 0.014 & 0.134 \\
		JP99 \cite{johnson1999} &
			{} & 1.10 & {} & {} &
			{--} & 0.062 & {} &
			0.656 & 0.0035 & 0.069 & 0.016 & 0.137 \\
		TO00 \cite{tomboulides2000} & 
			{} & {--} & {} & {} & 
			{--} & {--} & {} & 
			0.671 & 0.0028 & {--} & {--} & 0.136 \\
		KK01 \cite{kim2001} &
			{} & 1.087 & {} & {} &
			0.701 & 0.059 & {} &
			0.657 & {--} & 0.067 & {--} & 0.134 \\
		PW02 \cite{ploumhans2002} & 
			{} & {--} & {} & {} & 
			{--} & {--} & {} &
			0.683 & 0.0025 & 0.061 & 0.014 & 0.135 \\
		CS03 \cite{constantinescu2003} & 
			{} & {--} & {} & {} & 
			0.70 & 0.062 & {} & 
			0.655 & {--} & 0.065 & {--} & 0.136 \\
		WZ11 \cite{wang2011} & 
			{} & 1.13 & {} & {} & 
			{--} & {--} & {} & 
			0.68 & {--} & 0.071 & {--} & 0.135
	\end{tabular}
	\caption[%
    Drag and lift coefficients, and Strouhal numbers for a sphere at $\text{Re}=100,\,250,\,\text{and}\,\,500$.
  ]{%
    Drag and lift coefficients, and Strouhal numbers for a sphere at different Reynolds numbers.
    Present (A) and (B) correspond to test cases computed using $\Delta x / D \simeq 1.8 \times 10^{-2}$ and $\Delta x / D \simeq 8.8 \times 10^{-2}$, respectively.
    Results from JP99 -- \citet{johnson1999}, TO00 -- \citet{tomboulides2000}, KK01 -- \citet{kim2001}, PW02 -- \citet{ploumhans2002}, CS03 -- \citet{constantinescu2003}, and WZ11 -- \citet{wang2011} are also provided.%
    \label{tab:sphere_coeffs}
  }
\end{table}

The spread of values shown in Table~\ref{tab:sphere_coeffs} for spheres at $\text{Re}=300$ indicates that this test case is difficult to compute accurately.
The large spread of values for $\Delta C_\text{L}$ and $\Delta C_\text{D}$ (largest spread based on relative differences) has been attributed to differences in the domain size and boundary conditions of different numerical methods \cite{ploumhans2002}.
Consistent with this argument are the small differences in $\Delta C_\text{L}$ and $\Delta C_\text{D}$ between the present (B) results and those of the unbounded domain vortex method \cite{ploumhans2002}.

\begin{figure}[htbp]
  \begin{center}
    \includegraphics[width=0.95\textwidth]{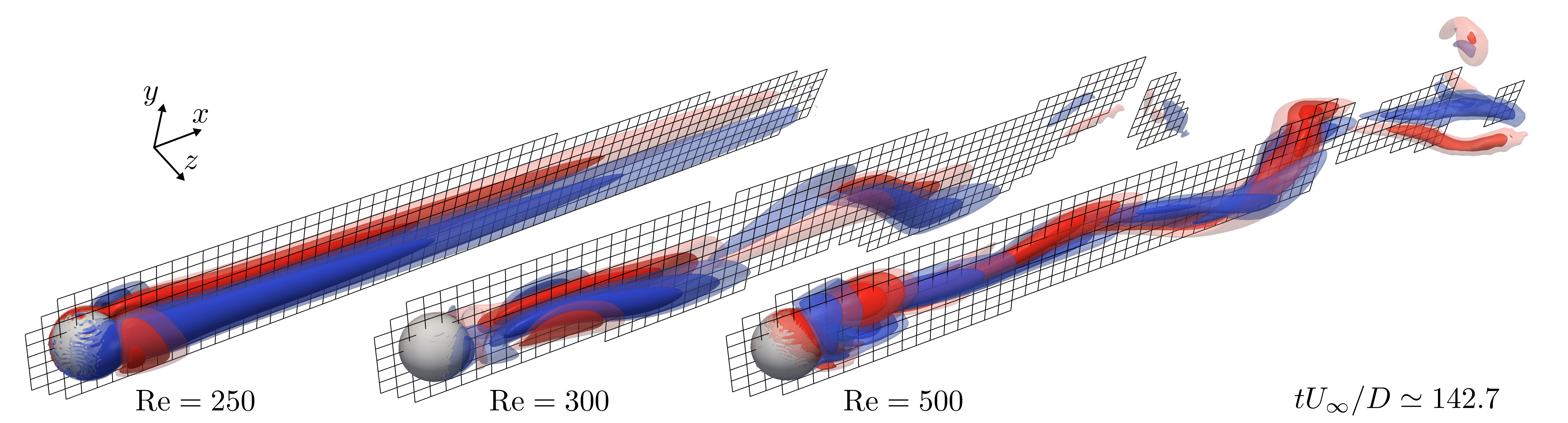}
  \end{center}
  \caption[%
    Stream\hyp{}wise vortices in the wake of a sphere at $\text{Re}=250,\,300,$ and $500$ depicted as iso\hyp{}surfaces of constant $\omega_x$.
  ]{%
    Stream\hyp{}wise ($x$-direction) vortices in the wake of spheres at different Reynolds numbers depicted as iso\hyp{}surfaces of constant $\omega_x$.
    Iso\hyp{}surfaces are for values of $\omega_x = \pm 0.20,\,0.10,\,\text{and}\,\, 0.05$ at $\text{Re}=250$, and of $\omega_x = \pm 1.0,\,0.50,\,\text{and}\,\, 0.25$ at $\text{Re}=300\,\,\text{and}\,\,500$.
    Depicted boxes are described in the caption of Figure~\ref{fig:verif_plate_highre}.
    \label{fig:sph_midre}%
  }
\end{figure}

The numerical experiments on spheres discussed thus far have considered the steady axis\hyp{}symmetric ($\text{Re}=100$), the steady planar\hyp{}symmetric ($\text{Re}=250$), and the periodic planar\hyp{}symmetric ($\text{Re}=300$) flow regimes.
Next we verify that the flow is unsteady with no fixed planar symmetry at $\text{Re} = 500$ \cite{tomboulides2000,ploumhans2002}.
Figure~\ref{fig:sph_midre} provides snapshots of stream\hyp{}wise vorticity iso\hyp{}surfaces for the aforementioned flow regimes (case of $\text{Re}=100$ is not shown since stream\hyp{}wise vorticity is of negligible magnitude).
The flow at $\text{Re} = 500$ is approximately symmetric about the $x$-$y$ plane at early times (similar to the $\text{Re}=300$ test case), but such symmetry is lost at later times as shown in Figure~\ref{fig:sph_midre} by the axial rotation of the stream\hyp{}wise vortices.

As a final demonstration of the IB\hyp{}LGF method, we consider the turbulent flow around a sphere at $\text{Re}=\numprint{3700}$.
This flow has been characterized in previous numerical \cite{seidl1998,yun2006,rodriguez2011} and experimental \cite{kim1988} investigations.%
\footnote{%
Previous investigations were conducted at $\text{Re}=\numprint{3700}$ \cite{yun2006,rodriguez2011}, $\text{Re}=\numprint{4200}$ \cite{kim1988},  and $\text{Re}=\numprint{5000}$ \cite{seidl1998}.}
The flow is computed for $0 \le t^* / U \le 60$ using $\numprint{81920}$ markers and $\Delta x \simeq 4.3 \times 10^{-3}$, where $t^*$ is used to indicate that flow was initialized from the large\hyp{}time solution of a sphere at $\text{Re}=\numprint{1000}$.
Subsequently reported time\hyp{}averaged values are computed over the last five large\hyp{}scale vortex shedding cycles ($\text{St}=0.215$ \cite{rodriguez2011}).

The thin boundary layer on the surface of the sphere is expected to be sufficiently well\hyp{}resolved since the present value of $(\Delta x)\text{Re}^{\frac{1}{2}}$ (scaling based on the expected $\mathcal{O}(\text{Re}^{-\frac{1}{2}})$ laminar boundary layer thickness \cite{schlichting2003}) is between the values of $(\Delta x)\text{Re}^{\frac{1}{2}}$ used to compute test cases (A) and (B) at $\text{Re}=300$.
As a point of reference, spheres at $\text{Re}=\numprint{3700}$ have been previously computed using a IB/LES method combined with a stretched grid with a near\hyp{}surface minimum spacings of $9 \times 10^{-3} D$ \cite{yun2006} and using a unstructured mesh with a near\hyp{}surface minimum element side lengths of $1.5 \times 10^{-3} D$ \cite{rodriguez2011}.
The \emph{a posteriori} grid analysis of \cite{rodriguez2011} demonstrates that the turbulent flow, with a minimum Kolmogorov length scale of $\eta/D = 1.34 \times 10^{-2}$ occurring in the $x/D < 3$ wake region, is well\hyp{}resolved by a second\hyp{}order unstructured mesh with an average element side length of $\overline{h}/D = 8 \times 10^{-3}$ over the $x/D < 3$ wake region.
Based on these grid considerations, we assume that the present set of grid parameters are adequate to capture both the thin laminar boundary layer on the surface and the turbulent wake of the flow.

\begin{figure}[htbp]
  \begin{center}
    \includegraphics[width=0.95\textwidth]{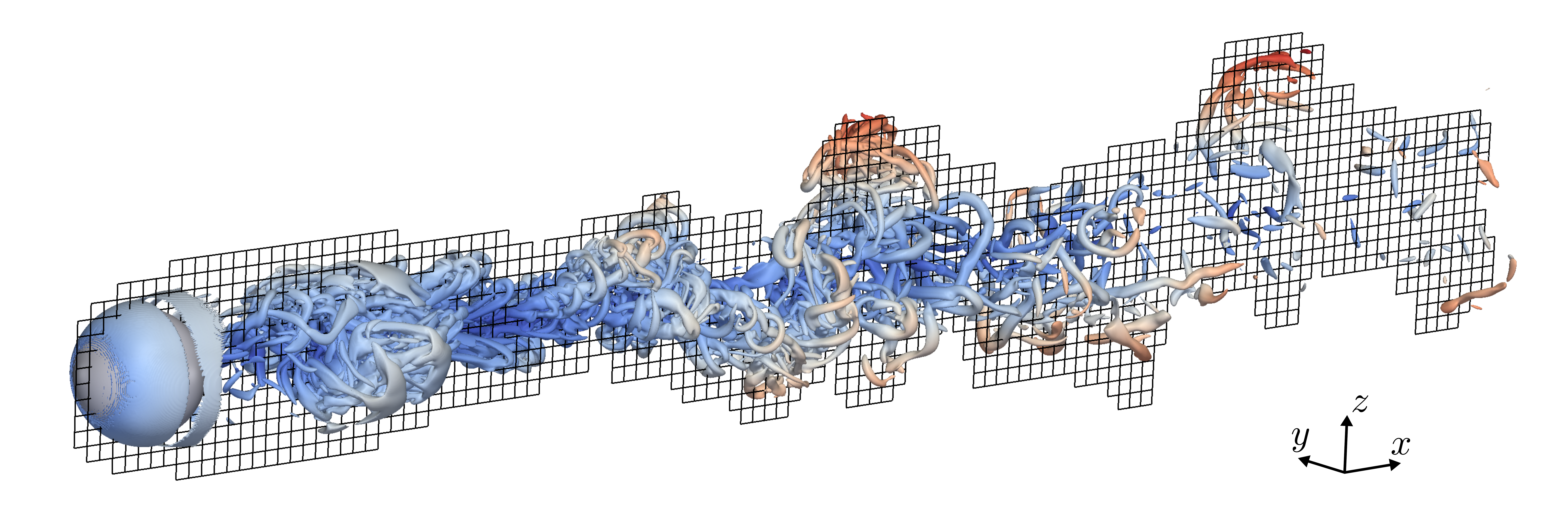}
  \end{center}
  \caption[%
    Vortex cores in the wake of a sphere at $\text{Re}=\numprint{3700}$ depicted as iso\hyp{}surfaces of constant $Q$-value.
  ]{%
    Vortex cores in the wake of a sphere at $\text{Re}=\numprint{3700}$ are illustrated by iso\hyp{}surfaces of constant $Q$-value.
    Depicted are iso\hyp{}surface of $Q D^2/U^2 = 2$ colored according to the radial distance from the center\hyp{}line of the sphere in the stream\hyp{}wise direction.
    Depicted boxes are described in the caption of Figure~\ref{fig:verif_plate_highre}.
    \label{fig:sph_highre_qcrit}%
  }
\end{figure}

The core of vortical structures in the wake are depicted as iso\hyp{}surfaces of constant $Q$-value in Figure~\ref{fig:sph_highre_qcrit}.
The $Q$-criterion \cite{hunt1988} defines coherent vorticies as connected regions where $Q$, the second invariant of $\nabla \mathbf{u}$, is positive.
Positive $Q$-values indicate a local excess of the rotation rate compared to the strain rate.
Figure~\ref{fig:sph_highre_qcrit} confirms the previously reported \cite{yun2006,rodriguez2011} pronounced helical\hyp{}like pattern of large\hyp{}scale vortical structures in wake.
A visual analysis of multiple snapshots verifies that the dominant vorticies forming the helical\hyp{}like pattern are convected downstream without significant axial rotations and that the pattern is the result of the apparent random azimuthal position of growing shear layer instabilities \cite{yun2006,rodriguez2011}.
We clarify that the strong small\hyp{}scale vortical filament- and horseshoe\hyp{}like structures in the downstream wake regions shown in Figure~\ref{fig:sph_highre_qcrit} are not readily seen in comparable plots of previous numerical experiments \cite{yun2006,rodriguez2011}, but this is expected from the fact that these previous experiments aggressively coarsen downstream grid regions.

We further characterize the flow by reporting on the large\hyp{}time mean surface stresses and net body forces.
The $i$-th component of the \emph{stress vector}, $\sigma_i = \left[ \mathbf{t} \right]_i$, at the location of the $q$-th IB marker can be approximated as $\sigma_i(\boldsymbol{\xi}_q) \approx \left[ \mathbf{f}_q \right]_i / A_q$, where $A_q$ is the area associated with the $q$-th IB marker.
Given that our IB markers are located at the centroids of the faces (triangles) of an icosphere, we take the $A_q$ to be equal to the area of the corresponding face.
The non\hyp{}dimensional surface stress coefficients in spherical coordinates are taken to be
\begin{equation}
  C_{\sigma,\mathbf{r}} = \frac{ \sigma_{r} - p_0 }{ \rho U },\quad
  C_{\sigma,\boldsymbol{\theta}} = \frac{ \sigma_{\theta} }{ \rho U },\quad
  C_{\sigma,\boldsymbol{\phi}} = \frac{ \sigma_{\phi} }{ \rho U },
  \label{eq:traction_coeffs}
\end{equation}
where $\mathbf{t} = \sigma_{r} \hat{\mathbf{r}} + \sigma_{\theta} \hat{\boldsymbol{\theta}} + \sigma_{\phi} \hat{\boldsymbol{\phi}}$, and $\theta$ and $\phi$ correspond to the polar and azimuthal angles, respectively (stagnation point located at $\theta = 0$).
Here, the reference pressure $p_0$ is taken to be $p_\infty - p_\text{sphere}$, where $p_\text{sphere}$ is the value of the approximately uniform pressure distribution inside the sphere.%
\footnote{%
The slight porosity of the numerical immersed surface results in an approximately uniform time\hyp{}dependent pressure distribution inside the sphere.
At $t U /D \approx 50$ the difference between the minimum and maximum pressure inside the sphere, but not in the support of $\mathscr{I}$, is approximately 0.3\% of $\frac{1}{2} \rho U^2$.
The maximum difference in $p_\text{sphere}$ between any two instantaneous measurements during $30 \le t U/D \le 60$ is approximately 1\% of $\frac{1}{2} \rho U^2$.
}
In the continuum limit, the surface normal stress coefficient $C_{\sigma,r}$ is equal to half of the pressure coefficient $C_p = ( p - p_\infty )/( \frac{1}{2} \rho U )$.
As discussed in Section~\ref{sec:verif_error}, raw point\hyp{}wise values of $\boldsymbol{\sigma}$ contain unphysical large high\hyp{}frequency oscillations.
These oscillations are partially filtered out and the point\hyp{}wise accuracy of $\boldsymbol{\sigma}$ is significantly improved using the boundary force post\hyp{}processing technique \cite{_goza2015}, which can be interpreted as a spatial weighted moving average smoothing technique that uses $\delta_{\Delta x}$ as the smoothing kernel.
This technique constructs smoothed boundary forces $\hat{\mathfrak{f}}$ by evaluating the expression $\hat{\mathfrak{f}} = \mathscr{I} \mathsf{W} \mathscr{I}^\dagger \mathfrak{f}$, where $\mathsf{W}(\mathbf{n})$ is equal to $1/\gamma(\mathbf{n})$ for the case of non\hyp{}zero $\gamma(\mathbf{n})=[\mathscr{I}^\dagger \mathbf{1}](\mathbf{n})$ and equal to zero otherwise.

\begin{figure}[htbp]
  \begin{center}
    \includegraphics[width=0.95\textwidth]{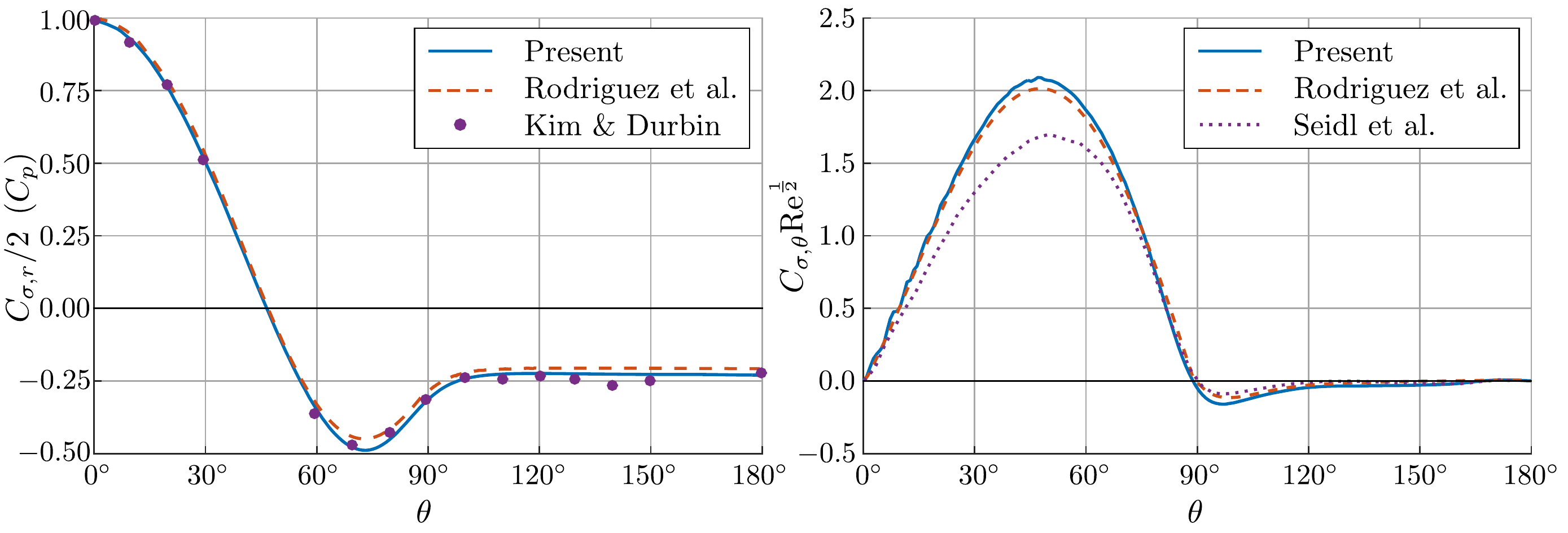}
  \end{center}
  \caption[%
    Time averaged pressure and skin\hyp{}friction coefficients as functions of the polar angle for a sphere at $\text{Re}=\numprint{3700}$.
  ]{%
     Time averaged pressure (\emph{left}) and skin\hyp{}friction (\emph{right}) coefficients as functions of the polar angle, $\theta$, for a sphere at $\text{Re}=\numprint{3700}$.
     Results compared to values reported by \citet{rodriguez2011} (DNS at $\text{Re}=\numprint{3700}$), \citet{kim1988} (exp. at $\text{Re}=\numprint{4200}$), and \citet{seidl1998} (DNS at $\text{Re}=\numprint{5000}$).
    \label{fig:sph_trac}%
  }
\end{figure}

Time\hyp{}averaged values of $C_p$ and $C_{\sigma,\boldsymbol{\theta}}$ as functions of the polar angle, $\theta$, are depicted in Figure~\ref{fig:sph_trac}.%
\footnote{%
The normalized skin\hyp{}friction coefficient, $C_{\sigma,\boldsymbol{\theta}} \text{Re}^{\frac{1}{2}}$, depicted in \cite{rodriguez2011} for the computations of \cite{seidl1998} are approximately 16\% larger than to those shown in the left plot of Figure~\ref{fig:sph_trac}.
The curve shown in Figure~\ref{fig:sph_trac} was computed by scaling the values of $\sigma_{\theta} \text{Re} / \rho U$ reported in \cite{seidl1998} by $\text{Re}^{-\frac{1}{2}}$, where $\text{Re}$ is taken to be the Reynolds number at which the numerical simulation was performed, i.e $\text{Re}=\numprint{5000}$.%
}
The present values are in good visual agreement with the body\hyp{}fitted mesh DNS values reported in \cite{rodriguez2011}.
We clarify that the curves shown in Figure~\ref{fig:sph_trac} include a small $\mathcal{O}(\Delta s)$ post\hyp{}processing error resulting from interpolating values of $\hat{\mathfrak{f}}$, which is defined on the faces of a six\hyp{}times subdivided icosphere, onto geodesic lines between $\theta=0^\circ$ and $\theta=180^\circ$.
Small remnants of the unphysical high\hyp{}frequency oscillations in $\mathfrak{f}$ are visually noticeable in the values of $C_{\sigma,\boldsymbol{\theta}}$ over the region of $0^\circ \le \theta \lesssim 50^\circ$.%
\footnote{%
Visual inspections three\hyp{}dimensional plots of the distribution of $\boldsymbol{\sigma}$ indicate that the oscillations $C_{\sigma,\boldsymbol{\theta}}$ for $0^\circ \le \theta \lesssim 50^\circ$ are in fact small oscillations in $\boldsymbol{\sigma}$ as opposed to oscillatory errors resulting from the interpolating values onto geodesic lines.}
Although these are undesirable features of the present non\hyp{}body\hyp{}conforming discretization, we find the magnitude of this error, $\Delta C_{\sigma,\boldsymbol{\theta}} \approx 0.1 \text{Re}^{-\frac{1}{2}} \approx 0.006$, to be acceptable.

Lastly, we report in Table~\ref{tab:sphere_highre} mean values for the drag coefficient ($\overline{C_\text{D}}$), base pressure coefficient ($\overline{C_{p}}{}_{,\text{b}}$), separation angle ($\overline{\theta_s}$), and polar locations of the minimum surface pressure ($\overline{\theta_{p}}{}_{,\text{min}}$) and of the maximum skin friction ($\overline{\theta_{\tau}}{}_{,\text{max}}$).
With the exception of $\overline{C_{p}}{}_{,\text{b}}$, the present values are within 2.1\% of those reported in \cite{rodriguez2011}. 
The difference in $\overline{C_{p}}{}_{,\text{b}}$ is also seen to be small, i.e. approximately 2.3\%, when compared to the maximum $\overline{C_{p}}$ shown in Figure~\ref{fig:sph_trac}.
The value of $\overline{C_\text{D}}$ reported in \cite{yun2006} (LES) is approximately 12\% lower than the $\overline{C_\text{D}}$ values reported here and in \cite{rodriguez2011} (DNS).
The grid refinement and mean turbulent statistics studies of \cite{rodriguez2011} attribute this discrepancy in $\overline{C_\text{D}}$ value to the sub\hyp{}grid model used in the numerical experiments of \cite{yun2006}.

\begin{table}[htbp]
	\centering
	\begin{tabular}{rcllllll}
		{} & {} &
			\multicolumn{1}{c}{$\text{Re}$} &
			\multicolumn{1}{c}{$\overline{C_\text{D}}$} &
			\multicolumn{1}{c}{$\overline{C_{p}}{}_{,\text{b}}$} &
      \multicolumn{1}{c}{$\overline{\theta_s}$} &
			\multicolumn{1}{c}{$\overline{\theta_{p}}{}_{,\text{min}}$} &
			\multicolumn{1}{c}{$\overline{\theta_{\tau}}{}_{,\text{max}}$} \\%
		\cline{3-8} \rule{0pt}{3ex}
    Present & DNS &
    	$\numprint{3700}$ &
      $0.389$ &
      $-0.230$ &
      $88.9^\circ$ &
      $73^\circ$ &
      $47^\circ$ \\
		YK06 \cite{yun2006} & LES &
			$\numprint{3700}$ &
      $0.355$ &
      $-0.194$ &
      $90^\circ$ &
      -- &
      -- \\
		RB11 \cite{rodriguez2011} & DNS &
			$\numprint{3700}$ &
      $0.393$ &
      $-0.207$ &
      $89.3^\circ$ &
      $72^\circ$ &
      $48^\circ$ \\
		KD88 \cite{kim1988} & exp. &
			$\numprint{4200}$ &
      -- &
      $-0.224$ &
      -- &
      -- &
      -- \\
		SM98 \cite{seidl1998} & DNS &
			$\numprint{5000}$ &
      $0.38$ &
      -- &
      $89.5^\circ$ &
      $71^\circ$ &
      $50^\circ$ \\
	\end{tabular}
	\caption[%
    Mean flow features of a sphere at Reynolds numbers between $\numprint{3700}$ and $\numprint{5000}$.
  ]{%
	Mean flow features of a sphere at Reynolds numbers between $\numprint{3700}$ and $\numprint{5000}$.
	Results from YK06 -- \citet{yun2006}, RB11 -- \citet{rodriguez2011}, KD88 -- \citet{kim1988}, and SM98 -- \citet{seidl1998} are provided.%
	\label{tab:sphere_highre}}
\end{table}

\section{Conclusions}
\label{sec:conclusions}

A computationally efficient IB method for external, viscous, incompressible flows around immersed surfaces with prescribed kinematics has been presented.
The IB\hyp{}LGF method is a significant extension of the LGF flow solver \cite{_liska2015}, which retains the efficiency and robustness of the flow solver by coupling a Lagrange multiplier treatment of the discrete boundary forces and the discretized no\hyp{}slip constraint with existing and new LGF techniques.
The semi\hyp{}discrete equations resulting from the formal spatial discretization of the incompressible Navier\hyp{}Stokes equations on unbounded staggered Cartesian grids and the discrete delta function treatment of the IB regularization and interpolation operators is shown to constitute a DAE system of index 2.
Using appropriately specialized order conditions for HERK integrators we proposed a few time integration schemes, which, when coupled with a viscous integrating factor technique, efficiently solve the discrete momentum ODE and the discrete divergence\hyp{}free and no\hyp{}slip constraints under a well\hyp{}understood theoretical framework.

Fast flow solutions are facilitated by using a projection\hyp{}like solver for the linear systems of equations arising from the implicit coupling the velocity, pressure, and boundary forces of the IF\hyp{}HERK scheme.
Unlike classical projection techniques, the present nested projection method is free of operator approximations, which in turn preserves the formal properties of the DAE time integration technique.
This method is equivalent to a LU decomposition of the linear system and is formulated as two sequential intermediate velocity and pressure computation steps, followed by a single boundary force solution step, and finalized by two sequential pressure and velocity correction steps.
Computational considerations for efficient iterative and direct boundary force solution techniques are discussed, and it is demonstrated that for many practical flows involving rigid surfaces a Cholesky\hyp{}based pre\hyp{}processing technique results in force solutions that require negligible computation times.
The pre\hyp{}processing technique results in a flow solver that depends on the solution of one additional discrete elliptic problem, i.e. force correction on the pressure, which is shown, by virtue of the flexible source and target regions of the LGF solver, to require significantly less computation time than the discrete pressure Poisson problem inherent to the flow solver (less than 50\% for the numerical experiments considered).

We implemented a parallel version of the IB\hyp{}LGF method for the case of rigid surfaces following the block\hyp{}wise adaptive grid of the LGF flow solver.
Modifications to the adaptivity criteria, grid sub\hyp{}domains, and parallel load balancing procedures were performed in order to efficiently and accurately capture the flow near immersed surfaces.
Detailed spatial and temporal refinement studies on flows around spheres were used to verify the expected convergence rates of the formulation.
Comparisons with previous numerical investigations on flows around rectangular flat plates and spheres at Reynolds numbers up to $\numprint{3700}$ were used to confirm the physical fidelity of computed solutions.
We also showed that accurate surface stresses can be obtained from the computed boundary forces using the post\hyp{}processing technique of \cite{_goza2015}.
All together, the present numerical experiments have demonstrated that the IB\hyp{}LGF method can overcome many of the limitations of previous IB methods including robust rigid\hyp{}surface solutions, accurate and efficient unbounded domain flow solutions, physical surface stress solutions, and the feasibility of fast, accurate numerical solutions to high (based on present day DNS capabilities) Reynolds numbers flows.

\section*{Acknowledgments}

This work was partially supported by the United States Air Force Office of Scientific Research (FA950--09--1--0189) and the Caltech Field Laboratory for Optimized Wind Energy with Prof. John Dabiri as PI under the support of the Gordon and Betty Moore Foundation.

\appendix

\section{Lattice Green's functions representations}
\label{app:lgfs}

The present formulation computes the action of $\mathsf{L}^{-1}$, $\mathsf{E}(\alpha)$, and $\mathsf{K}(\alpha)$ as discrete convolutions, e.g. Eq~\eqref{eq:dpoisson_conv}, of $\mathsf{G}_{\mathsf{L}}$, $\mathsf{G}_{\mathsf{E}}(\alpha)$, and $\mathsf{G}_{\mathsf{K}}(\alpha)$.
Expressions for these LGFs in terms of Fourier and Bessel integrals are given by
\begin{subequations}
  \begin{align}
    [\mathsf{G}_{\mathsf{E}}(\alpha)](\mathbf{n})
      &= \frac{1}{8\pi^3}
        \int_{\Pi} e^{ -i \mathbf{n}\cdot\boldsymbol{\xi} - \sigma(\boldsymbol{\xi}) } 
        \,d \boldsymbol{\xi}
      = \prod_{q\in\mathbf{n}} \left[ e^{-2\alpha} I_{q}(2\alpha) \right],
      \label{eq:lgf_intfac} \\
    (\Delta x)^2 \mathsf{G}_{\mathsf{L}}(\mathbf{n})
      &= \frac{1}{8\pi^3}
        \int_{\Pi} \frac{ e^{ -i \mathbf{n}\cdot\boldsymbol{\xi} } }
        	{ \sigma(\boldsymbol{\xi}) } \, d\boldsymbol{\xi}
      = - \int_{0}^{\infty} [\mathsf{G}_{\mathsf{E}}(t)](\mathbf{n}) \, dt,
      \label{eq:lgf_lap} \\
    (\Delta x)^2 [\mathsf{G}_{\mathsf{K}}(\alpha)](\mathbf{n})
    	&= \frac{1}{8\pi^3}
        \int_{\Pi} \frac{ e^{ -i \mathbf{n}\cdot\boldsymbol{\xi} - \sigma(\boldsymbol{\xi}) } }
          { \sigma(\boldsymbol{\xi}) } \, d\boldsymbol{\xi}
      = - \int_{\alpha}^{\infty} [\mathsf{G}_{\mathsf{E}}(t)](\mathbf{n}) \, dt,
      \label{eq:lgf_iflap}
  \end{align}
  \label{eq:lgfs}%
\end{subequations}
where $\sigma(\boldsymbol{\xi}) = 2\cos(\xi_1) + 2\cos(\xi_2) + 2\cos(\xi_3) - 6$, $\Pi=(-\pi,\pi)^3$, and $I_n(z)$ is the modified Bessel function of the first kind of order $n$.

Here, we introduce a simple procedure for efficiently computing $\left[\mathsf{G}_{\mathsf{K}}(\alpha)\right](\mathbf{n})$ and refer the reader to the discussions of \cite{_liska2014} and \cite{_liska2015} for examples of numerical techniques used to evaluate $\mathsf{G}_{\mathsf{L}}(\mathbf{n})$ and $[\mathsf{G}_{\mathsf{E}}(t)](\mathbf{n})$.
We consider the partition of $\left[\mathsf{G}_{\mathsf{K}}(\alpha)\right](\mathbf{n})$ given by
\begin{equation}
  \left[\mathsf{G}_{\mathsf{K}}(\alpha)\right](\mathbf{n}) 
    = \mathsf{G}_{\mathsf{L}}(\mathbf{n}) + \left[\mathsf{R}(\alpha)\right](\mathbf{n}),
\end{equation}
where $\left[\mathsf{R}(\alpha)\right](\mathbf{n}) = (\Delta x)^{-2} \int_{0}^{\alpha} [\mathsf{G}_{\mathsf{E}}(t)](\mathbf{n}) \, dt$.
The combined look\hyp{}up table and asymptotic expansion approach of \cite{_liska2014} is used to compute the first term, $\mathsf{G}_{\mathsf{L}}(\mathbf{n})$, and an adaptive Gauss\hyp{}Kronrod (GK) integration scheme is used to evaluate the second term, $\left[\mathsf{R}(\alpha)\right](\mathbf{n})$.
For large values of $\mathbf{n}$ few, if any, subdivisions are required by the GK scheme since the value of $\left[\mathsf{R}(\alpha)\right](\mathbf{n})$ is significantly smaller than the value of $\mathsf{G}_{\mathsf{L}}(\mathbf{n})$.%
\footnote{%
The leading order term in the asymptotic expansion of $\mathsf{G}_{\mathsf{L}}(\mathbf{n})$ is $1/(4\pi|\mathbf{n}|)$ \cite{_liska2014}.
For a fixed $\alpha$, the integrand $\left[\mathsf{R}(\alpha)\right](\mathbf{n})$, i.e. $[\mathsf{G}_{\mathsf{E}}(t)](\mathbf{n})$, decays faster than any exponential \cite{_liska2015}.}
Lastly, we note that evaluating discrete convolution of LGFs using the LGF\hyp{}FMM \cite{_liska2014} only requires the point\hyp{}wise values of LGFs to be computed once, as a pre\hyp{}processing step, per simulation.

\section{Half\hyp{}explicit Runge\hyp{}Kutta schemes}
\label{app:herks}

The IF\hyp{}HERK schemes used to perform the numerical experiments of Section~\ref{sec:verif} are:
\begin{gather}
  \stackrel{\mbox{\small{\textsc{Scheme A}}}}{%
    \begin{array}{c|ccc}
      \multicolumn{1}{r}{{}} & {} & {} & {} \\
      0 & {} & {} & {} \\
      \textstyle\frac{1}{2} & \textstyle\frac{1}{2} & {} & {} \\
      1 & \textstyle\frac{\sqrt{3}}{3} & \textstyle\frac{3-\sqrt{3}}{3} & {} \\
      \hline
      {} & \textstyle\frac{3+\sqrt{3}}{6} & \textstyle\frac{-\sqrt{3}}{3} & \textstyle\frac{3+\sqrt{3}}{6}
    \end{array} }
  \quad
  \stackrel{\mbox{\small{\textsc{Scheme B}}}}{%
  \begin{array}{c|ccc}
    \multicolumn{1}{r}{{}} & {} & {} & {} \\
    0 & {} & {} & {} \\
    \textstyle\frac{1}{3} & \textstyle\frac{1}{3} & {} & {} \\
    1 & -1 & 2 & {} \\
    \hline
    {} & 0 & \textstyle\frac{3}{4} & \textstyle\frac{1}{4}
  \end{array} }
  \quad
  \stackrel{\mbox{\small{\textsc{Scheme C}}}}{%
  \begin{array}{c|ccc}
    \multicolumn{1}{r}{{}} & {} & {} & {} \\
    0 & {} & {} & {} \\
    \textstyle\frac{8}{15} & \textstyle\frac{8}{15} & {} & {} \\
    \textstyle\frac{2}{3} & \textstyle\frac{1}{4} & \textstyle\frac{5}{12} & {} \\
    \hline
    {} & \textstyle\frac{1}{4} & 0 & \textstyle\frac{3}{4}
  \end{array} }
  \quad
  \stackrel{\mbox{\small{\textsc{Scheme D}}}}{%
  \begin{array}{c|cccc}
    0 & {} & {} & {} & {} \\
    \textstyle\frac{1}{2} & \textstyle\frac{1}{2} & {} & {} & {} \\
    \textstyle\frac{1}{2} & 0 & \textstyle\frac{1}{2} & {} & {} \\
    1 & \textstyle\frac{1}{4} & 0 & 0 & 1 \\
    \hline
    {} & \textstyle\frac{1}{6} & \textstyle\frac{1}{3} & \textstyle\frac{1}{3} & \textstyle\frac{1}{6}
  \end{array} }
  \label{eq:ifherk_schemes}
\end{gather}
The expected order of accuracy for Schemes A--D based on the simplified HERK order\hyp{}conditions discussed in Section~\ref{sec:discrete_time} are included in Table~\ref{tab:ifherk_schemes}.
As a point of comparison, Table~\ref{tab:ifherk_schemes} also includes the expected order of accuracy for problems with no immersed surfaces (i.e. Eq.~\eqref{eq:ibdnstp_trans:mom} and \eqref{eq:ibdnstp_trans:cont} with $\mathfrak{f}=0$) and for general semi\hyp{}explicit DAEs of index 2 (i.e. Eq.~\eqref{eq:dae_brasey}).

\begin{table}[htbp]
  \centering
  \begin{tabular}{c|ccccccc}
    {Scheme} & \emph{$y$} & \emph{$z$} & \emph{$y$}${}^+$ & \emph{$z$}${}^+$  & \emph{$y$}${}^*$ & \emph{$z$}${}^*$ & \\
    \cline{1-7}
    A & 2 & 2 & 2 & 2 & 2 & 2 \\
    B & 3 & 2 & 3 & 2 & 3 & 2 \\
    C & 2 & 1 & 3 & 1 & 2 & 1 \\
    D & 3 & 1 & 4 & 1 & 2 & 1
  \end{tabular}
  \caption[%
    Expected order of accuracy of IF-HERK schemes for the solution variable $y$ (velocity) and for the constraint variable $z$ (pressure and body forces.
  ]{%
    Expected order of accuracy of HERK schemes for the solution variable $y$ (velocity) and for the constraint variable $z$ (pressure and body forces).
    The superscripts $+$ and $*$ denotes values for problems with no immersed surface and for general semi\hyp{}explicit DAEs of index 2, respectively.
    \label{tab:ifherk_schemes}%
  }
\end{table}

Scheme B is the only three\hyp{}stage scheme with a third\hyp{}order accurate solution variable for general semi\hyp{}explicit DAEs of index 2 \cite{brasey1993}.
The RK coefficients of Scheme C and Scheme D correspond to the popular three\hyp{}stage fractional step method of \cite{le1991} and the four\hyp{}stage ``original'' RK method, respectively.
As discussed in \cite{_liska2015}, Scheme A has the advantage of having equispaced RK nodes, i.e. $c_i$'s, which reduce the number of distinct integrating factors.
Fewer distinct integrating factors reduces the number of pre\hyp{}processing operations and lowers the storage requirements of the LGF\hyp{}FMM \cite{_liska2014} and of the Cholesky\hyp{}based force Schur complement technique discussed in Section~\ref{sec:linsys_schur}.
Scheme A and D only require two distinct integrating factors (with one of them being the identity operator), as opposed to the three distinct integrating factors required by Scheme B and C.

In the absence of an immersed surface, a linear stability analysis about a uniform base flow $\mathbf{U}$ of the IF-HERK method \cite{_liska2015} indicates that solutions are subject to the CFL condition
\begin{equation}
  \text{CFL} = \frac{|\mathbf{U}| \Delta t}{\Delta x} < \text{CFL}_{\text{max}},
\end{equation}
where $\text{CFL}_\text{max}$ depends on the RK coefficients of the scheme.
The value of $\text{CFL}_\text{max}$ is unity for Schemes A--C and $\frac{2\sqrt{2}}{\sqrt{3}}$ for Scheme D.
In practice, we expect solutions to the non-linear governing equations to remain stable as long as the CFL conditions resulting from linearizing the flow at each grid point are satisfied, i.e. as long as $\max(|\mathbf{u}|) \Delta t / \Delta x < \text{CFL}_{\text{max}}$.

The CFL condition $\Delta x \sim \Delta t$ and the second\hyp{}order accuracy (in the absence of immersed surfaces) of the present solver imply that the potential reduction in the operation count resulting from higher than second\hyp{}order HERK schemes is limited.
As a result, the lower pre\hyp{}processing cost of Scheme A compared to Scheme C makes Scheme A the preferred HERK scheme for the present formulation.
Here, we did not consider Schemes C and D to be potential ``preferred'' schemes since they are only first\hyp{}order accurate in the constraint variables.

\bibliographystyle{model1-num-names}
\bibliography{jcp_iblgf}

\end{document}